\documentclass[lettersize,journal]{IEEEtran}
\usepackage{amsmath,amsfonts}
\usepackage{algorithm}
\usepackage{algpseudocode}
\usepackage{array}
\usepackage{textcomp}
\usepackage{stfloats}
\usepackage{url}
\usepackage{verbatim}
\usepackage{graphicx}
\usepackage{cite}
\usepackage{subcaption}
\usepackage{booktabs}
\usepackage{longtable}
\usepackage[table]{xcolor}

\usepackage{multirow}
\usepackage{multicol}

\newtheorem{theorem}{Theorem}[section]

\newtheorem{definition}[theorem]{Definition}
\newtheorem{assumption}[theorem]{Assumption}

\begin{document}
\title{QuantumShield: Multilayer Fortification for Quantum Federated  Learning}
\author{Dev Gurung and Shiva Raj Pokhrel
\thanks{Authors are with School of IT, Deakin University, Australia; Corresponding Author: shiva.pokhrel@deakin.edu.au}
\thanks{Manuscript received April 19, 2021; revised August 16, 2021.}}
\markboth{Journal of \LaTeX\ Class Files,~Vol.~14, No.~8, August~2021}
{Shell \MakeLowercase{\textit{et al.}}: A Sample Article Using IEEEtran.cls for IEEE Journals}

\maketitle

\begin{abstract}
In this paper, we propose a groundbreaking quantum-secure federated learning (QFL) framework designed to safeguard distributed learning systems against the emerging threat of quantum-enabled adversaries. 
As classical cryptographic methods become increasingly vulnerable to quantum attacks, our framework establishes a resilient security architecture that remains robust even in the presence of quantum-capable attackers. 
We integrate and rigorously evaluate advanced quantum and post-quantum protocols
including Quantum Key Distribution (QKD), Quantum Teleportation, Key Encapsulation Mechanisms (KEM) and Post-Quantum Cryptography (PQC)
to fortify the QFL process against both classical and quantum threats. 
These mechanisms are systematically analyzed and implemented to demonstrate their seamless interoperability within a secure and scalable QFL ecosystem.
Through comprehensive theoretical modeling and experimental validation, this work provides a detailed security and performance assessment of the proposed framework. 
Our findings lay a strong foundation for next-generation federated learning systems that are inherently secure in the quantum era.
\end{abstract}


\section{Introduction}
Quantum computing is expected to significantly influence today's computing world by providing an opportunity for revolutionary advancements while also posing monumental security risks to current cryptographic systems \cite{bennettQuantumCryptographyPublic2014}.
Especially quantum computers present a significant challenge to asymmetric encryption and digital signature schemes based on number theory (such as RSA, DSA etc.), which are fundamental to the security of today's digital infrastructures \cite{Falconb}. 
The problem with classical cryptographic methods is that the mathematical problems on which they are based can be effectively compromised by quantum algorithms \cite{daiGOLFUnleashingGPUDriven2025}. 
These threats will then further add potential risks to distributed machine learning scenarios in frameworks like QFL, 
when access to operational quantum computers becomes more feasible.

In any distributed digital infrastructure, security is paramount in forming the fundamental basis for ensuring its reliability and trustworthiness.
It becomes even more critical due to the increased number of stakeholders involved in security matters in systems such as FL.
In the quantum era, threats may originate from either classical or quantum models.
Today we rely on classical cryptographic methods that are based on mathematical problems that are susceptible to being broken or resolved by quantum algorithms, 
making quantum computers a significant threat, as they can efficiently run such algorithms \cite{elliottQuantumCryptography2004}.
Various existing solutions to this threat that can be implemented and available are quantum cryptography \cite{elliottQuantumCryptography2004}, post-quantum cryptography \cite{schneier_nists_2022}, etc. 

\begin{figure}
    \centering
    \includegraphics[width=0.7\linewidth]{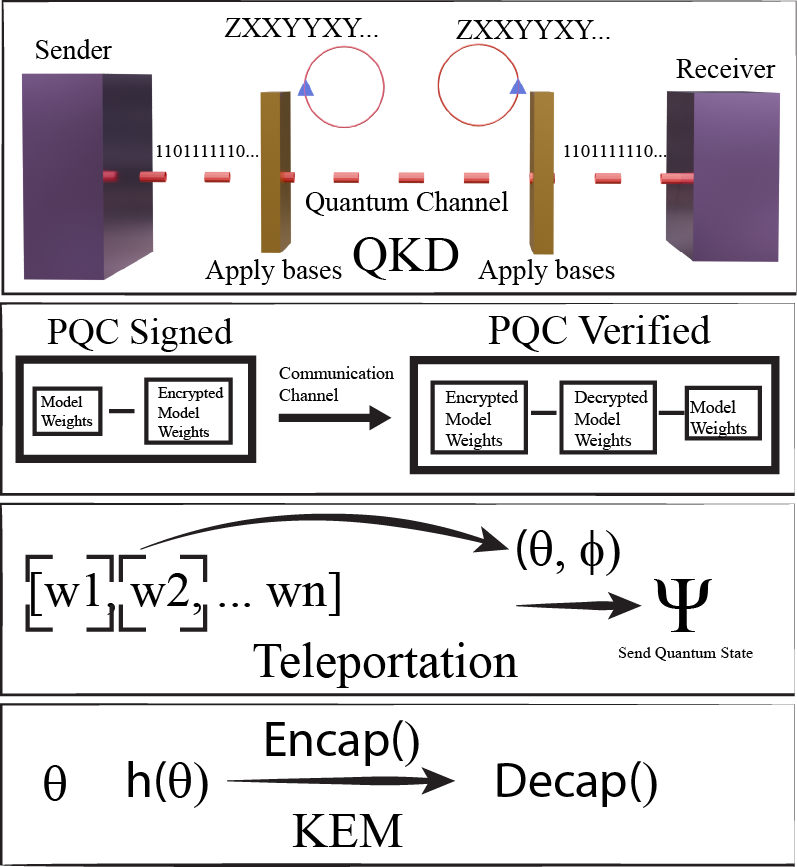}
    \caption{Pragmatic view of the proposed QuantumShield: Developing multi-Layer cryptographic fortification for QFL}
    \label{fig:secureQF_preview}
\end{figure}

For systems like QFL, numerous potential security vulnerabilities may arise.
It is thus crucial to ensure security and address risks in various aspects such as aggregation, communication, etc. to mention some.
For that, protocols must be carefully designed to implement additional measures such as preventing the central server from obtaining information or exerting influence on the local model of the client during the aggregation phase \cite{jereTaxonomyAttacksFederated2021}.
Compromised clients and any adversary can analyze global models to infer about the other client's dataset and affect the model performance by sending a poisoned model to the server.

Among many other threats, during the training phase, backdoor or Trojan attacks can aim to alter the model and learn unwanted misclassifications of the task \cite{jereTaxonomyAttacksFederated2021}.
Model poisoning can occur through gradient manipulation or training rule manipulation \cite{jereTaxonomyAttacksFederated2021}.
With gradient manipulation, adversaries can manipulate local model gradients.
Through training rule manipulation, adversaries can add a penalty term to the objective function while minimizing the distance between faulty and accurate updates.
Some threats, such as poisoning attacks in FL, include label flipping and the insertion of a backdoor \cite{jereTaxonomyAttacksFederated2021}.
With label flipping, adversaries aim to permute labels, while the features remain intact.
Unlike backdoor insertion, adversaries alter the training data itself.

Limited studies have been conducted on secure QFL, covering quantum-secure communication models specific to the QFL framework \cite{gurungSECURECOMMUNICATIONMODEL2023}, 
the integration of quantum-secure blockchain within FL networks \cite{gurungPerformanceAnalysisDesign2025} etc.
For classical FL, the works include a lattice-based federated learning protocol \cite{xuLaFLatticeBasedCommunicationEfficient2022e},  
a lightweight and secure federated learning solution, the LSFL scheme, offered by Zhang et al. \cite{zhangLSFLLightweightSecure2023}, 
while tackling the challenge of non-IID data has been addressed by Miao et al. \cite{miaoSecureModelContrastiveFederated2023}. 
With a focus on privacy-preserving federated learning schemes, secure weighted aggregation methods was proposed by He et al. \cite{heSecureWeightedAggregation2025}. 
In our work, we deviate from the existing literature by providing a comprehensive and meticulous approach to advance the security of the QFL framework specifically for the quantum era.

Although traditional security features, including those in classical FL, have been extensively studied, 
the investigation of quantum security related to protocols like QKD, quantum teleportation, KEMs, and PQC schemes, particularly for QFL contexts, remains sparse.
In addition, their implementation requires deeper study on their practicality and suitability, since they have their own limitations and challenges.
For example, the QKD protocol key may have limitations on the number of key sizes it can create \cite{metgerSecurityQuantumKey2023} and teleportation might be difficult to implement in networks like QFL \cite{hermansQubitTeleportationNonneighbouring}.
Similarly, we do not know how QKD works with classical encryption, Teleportation, in distribution environment like QFL, may need multiple sender receiver pair that demand multi-tier entanglement in multiple pairs, which can be hard to manage and implement.
In this study, our objective is to address these challenges along with various security challenges by investigating and developing multiple strategies to mitigate threats unique to the QFL framework. 
Given that QFL fundamentally differs from classical FL, we introduce innovative methodologies to execute and address inevitable security vulnerabilities in the QFL context.
In summary, the key contributions of this work are as follows.
\begin{enumerate}
    \item We present a secure quantum framework for quantum Federated Learning (QFL), developed using post-quantum cryptographic (PQCs) methodologies, quantum key distribution (QKD), teleportation, and key encapsulation mechanisms (KEMs). This framework facilitates quantum-encrypted communication to transmit trained and updated models between the central server and local clients.
    Our proposed framework is secure at the multilevel, such as in the local training phase, during the communication phase, etc.
    \item Through comprehensive theoretical and experimental scrutiny, we address the difficulties associated with implementing and designing a quantum secure QFL framework and the practicality of the proposed framework.
\end{enumerate}

\section{Background}
\subsection{Post Quantum Cryptography}
The design of various post-quantum cryptography is based on various techniques.
For example, one of the PQC schemes called
Dilithium is based on the ``Fiat-Shamir with Aborts" technique of Lyubashevsky \cite{alagicRecommendationsKeyEncapsulationMechanisms2025}.
It uses rejection sampling approach to make lattice-based Fiat-Shamir schemes secure.
\textit{CRYSTALS-Dilithium}\footnote{https://pq-crystals.org/dilithium/} is a digital signature scheme that is highly secure under chosen message attacks based on the hardness of lattice problems over module lattices.
This means producing a signature of a message whose signature hasn't been known is impossible. Similarly, 
producing a different signature for a message is not feasible even though the advisory has already seen is the signature.
\textit{CRYSTAL-Kyber}\footnote{https://pq-crystals.org/kyber/index.shtml} , another scheme, is an IND-CCA2-secure key encapsulation mechanism (KEM).
Its security is based on the difficulty in solving the Learning-with-Errors (LWE) problem on module lattices.
Others include
SPHINCS+\footnote{https://sphincs.org/} which is a stateless hash-based signature scheme and 
FALCON\footnote{https://falcon-sign.info/} which is based on lattice-based problems.


\subsection{Quantum Key Distribution}
QKD is used to share random bits through a communication channel without prior sharing of secret information \cite{bennettQuantumCryptographyPublic2014}.
While users consult through the use of ordinary non-quantum channel, which is subject to passive eavesdropping.
They can move on with agreeing or disagreeing to use the secret key based on whether there has been a disturbance during transmission.
QKD is based on fundamental concepts of quantum mechanics, which are Heisenberg's uncertainty principle, the no-cloning theorem, and quantum entanglement.
In summary, Heisenberg's uncertainty principle states that a pair of physical properties such as position and momentum cannot be measured simultaneously. That means we cannot know both the position and the speed of a particle such as a photon or electron with high accuracy, which can be mathematically presented as 
\[
\Delta x \Delta p \geq \frac{h}{4 \pi}
\]
where, $h$ is Planck's constant, $\Delta x$ and $\Delta p$ is uncertainty in position and momentum, respectively.
Quantum cryptography utilizes the polarization of photons on different bases as conjugate properties.
With no-cloning theorem, an Eve cannot make a perfect copy of an unknown quantum state.
With quantum entanglement, two quantum particles can be entangled with each other. 

\subsection{Teleportation}
Teleportation is a protocol in which a sender transmits a qubit to the receiver using a shared entangled quantum state as well as two bits of classical information \cite{brassard_teleportation_1998}.
Suppose that our aim is to send a qubit $Q$ in the state $|\psi\rangle$ from the sender to the receiver;
the prerequisite for this is that both the sender $A$ and receiver $B$ should each have a part of the entangled qubit pair $(A,B)$.
Now, to send the qubit $Q$, the sender performs a controlled NOT operation on pair $(A,Q)$ with $Q$ being the control bit and $A$ being the target, which is followed by the Hadamard gate operation on Q.
The sender then measures both the qubits $A$ and $Q$.
If we assume $a$ as the measurement output of qubit $A$ and $b$ as the measurement output of qubit $Q$, then, depending on the values of $a$ and $b$, after receiving $a$ and $b$ through the classical channel, the receiver performs the X or Z gate on qubit $B$.
After this operation, the qubit state $B$ is transformed into the original qubit state of $Q$ sent by the sender, i.e. $|\psi\rangle$.




\begin{figure*}
    \centering
    \includegraphics[width=\linewidth]{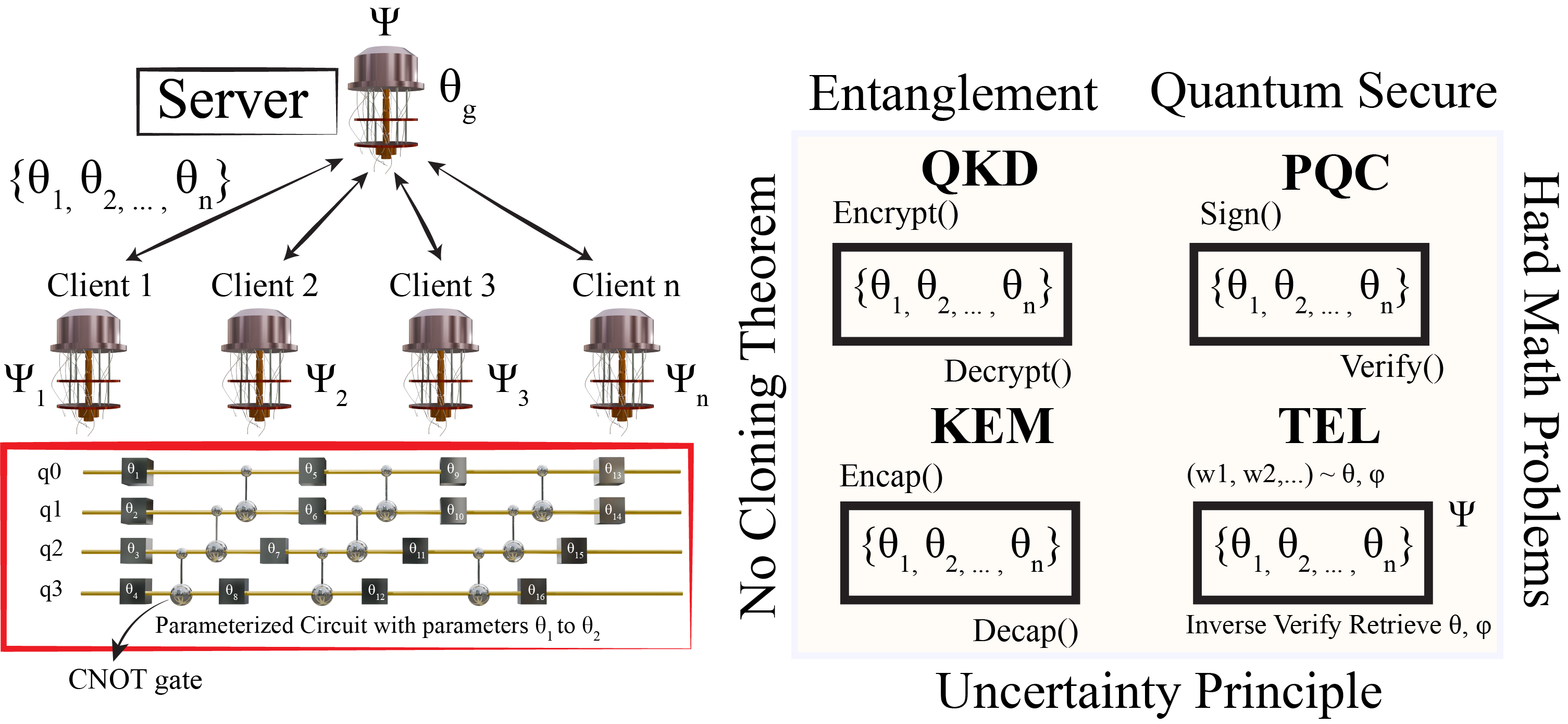}
    \caption{Secure QFL framework empowered by QKD, Teleportation, KEM and PQC schemes}
    \label{fig:secureQFL_framework}
\end{figure*}

\section{Related Work}
To this end, most of the existing work mostly centers around classical FL with almost nonexistent research in the field of secure QFL.

Early work on secure QFL includes the work by Gurung et al.\cite{gurungSECURECOMMUNICATIONMODEL2023} that proposed a secure communication model for the QFL framework and toward classical FL with a quantum-secure blockchain FL network \cite{gurungPerformanceAnalysisDesign2025}.
Both works involved some form of use of PQC protocols.
Xu et al. \cite{xuLaFLatticeBasedCommunicationEfficient2022e} proposed LaF, a lattice-based FL protocol with postquantum security. 
Using a multi-use secret scheme, LaF reduces communication overhead compared to Google's double-masking approach, improving runtime efficiency.

Zhang et al. \cite{zhangLSFLLightweightSecure2023} introduced LSFL, a lightweight and secure FL scheme for edge computing.
It uses a two-server secure aggregation protocol to ensure Byzantine robustness and privacy, 
maintaining model accuracy comparable to that of FedAvg, reducing computational overhead.
To mitigate model inversion attacks, Issa et al. \cite{issaRVEPFLRobustVariational2024} proposed RVE-PFL, a robust personalized FL approach based on variational encoders.
The method comprises a personalized variational encoder architecture and a trustworthy thread-model integrated FL approach which autonomously preserves data privacy and mitigates MI attacks.
The main challenges in FL such as the identification of impersonators in the participants, the implementation of efficient and secure privacy preservation methods, and the significant communication cost are addressed by Chen et al. \cite{chenSecureEfficientFederated2024}.
The work proposed an approach to combine advantages of secret sharing, Diffie-Hellman key agreement, function encryption to develop a verifiable secure multiparty computing algorithm.
The proposed algorithm, termed ESAFL, is an efficient, secure, and authenticated FL algorithm that leverages compressed sensing and an all-or-nothing transform to reduce communication costs and protect local gradients.

Other works include secure FL framework using stochastic quantization technique to balance communication efficiency, model accuracy and data privacy \cite{lyuSecureEfficientFederated2024}, 
robust and fair model aggregation solution Romoa-FL for cross-silo FL against poisoning attacks \cite{maoSecureModelAggregation2024}, 
secure FL scheme using model-constrastive loss and improved compressive sensing to address non-IID data challenge \cite{miaoSecureModelContrastiveFederated2023}, 
privacy preserving FL scheme with secure weighted aggregation \cite{heSecureWeightedAggregation2025}, 
VERIFL, communication efficient and fast verifiable aggregation protocol for FL \cite{guoVeriFLCommunicationEfficientFast2021}, multi qubit broadcast based QFL \cite{zhangEfficientSecureMultiQubit2025} etc.

The majority of existing studies utilize classical methods for cryptographic protocols, 
which are susceptible to quantum algorithms such as Shor's algorithm. 
Additionally, most research is concentrated on classical federated learning, 
fundamentally distinct from our study in distributed or federated quantum machine learning, 
known as QFL. 
Therefore, this study focuses on filling the literature gap in the field of QFL,
offering numerous strategies to counter quantum threats and ensuring compatibility with the quantum paradigm.

\section{Proposed Quantum Secure QFL}
In this section, we discuss the proposed protocols to secure the QFL framework.

\subsection{QKD QFL}
We propose an implementation protocol for quantum cryptography with QKD along with symmetric encryption techniques like One-Time Pad (OTP) or Fernet encryption methods.
QKD protocol used is BB84 \cite{bennettQuantumCryptographyPublic2014}.
Once both the client and server share the private key using the QKD protocol, we then utilize encryption protocols such as symmetric key encryptions.
Once each client completes local training, the
QKD private key is used to encrypt the model parameters and then sent to the server.
The server also has a private key associated with that particular client, and thus uses it to decrypt the model parameters.
After that, the server performs the standard FedAvg procedure to compute the global model.

\begin{figure}
    \centering
    \includegraphics[width=0.9\linewidth]{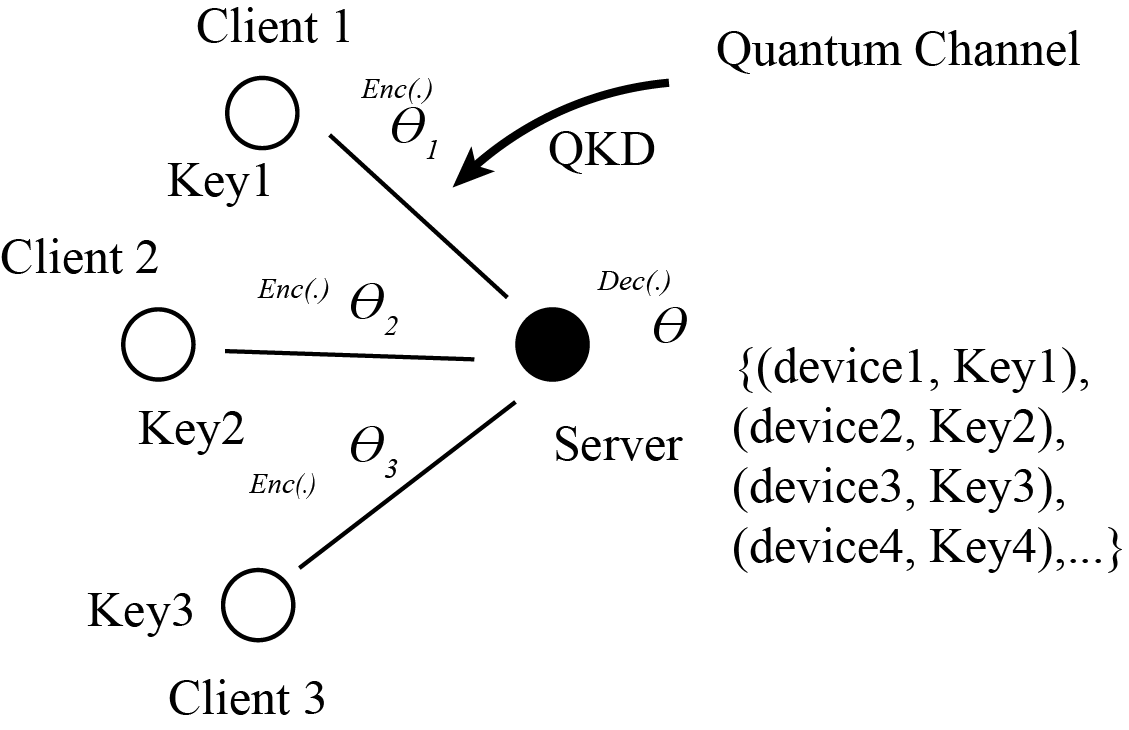}
    \caption{QKD QFL}
    \label{fig:qkd_qfl}
\end{figure}

With the QKD secure approach, the BB84 protocol \cite{bennettQuantumCryptographyPublic2014} is followed to ensure the secure transmission of the model parameters between local devices and the server.
As in Algorithm \ref{alg:qkd} and also depicted in Figure \ref{fig:qkd_qfl}, in each communication round, 
$n$ devices train the local VQC model to obtain parameters $\theta_i$.
Each device generates a QKD key pair $(Akey_i, Bkey_i)$ using quantum circuits with Hadamard gates and measurements to create shared secret keys. 
The parameters $\theta_i$ are encrypted using the Fernet or OTP approach with $Akey_i$ and sent to the server. 
The server decrypts the encrypted parameters using the corresponding
$Bkey_i$, aggregates them by computing the global parameters $\theta_g = \frac{1}{n} \sum_{i=1}^n \theta_i$, and encrypts $\theta_g$ with $Akey_i^s$ for each device.
The encrypted global parameters are distributed back to the devices that decrypt them using $Bkey_i^s$ to update their local VQCs.


\begin{algorithm}
\caption{QKD QFL}
\label{alg:qkd}
\begin{algorithmic}[1]
\State \textbf{Input}: $n$ devices, VQC parameters $\theta_i$
\State Initialize local VQC models with random $\theta_i$
\For{each round $t = 1$ to $T$}
    \For{each device $i = 1$ to $n$}
        \State Train local VQC to update $\theta_i^{(t)}$
        \State Generate QKD keys $(AKey_i, BKey_i)$ via BB84
        \State Encrypt $\theta_i^{(t)}$ with $Key_i^A$ to get $Enc(\theta_i^{(t)})$
        \State Send $Enc(\theta_i^{(t)})$ to server
    \EndFor
    \State Server decrypts $\{Enc(\theta_i^{(t)})\}$ using $\{BKey_i\}$
    \State Compute global $\theta_g^{(t)} = \frac{1}{n} \sum_{i=1}^n \theta_i^{(t)}$
    \For{each device $i = 1$ to $n$}
        \State Generate QKD keys $(AKey_i^{s}, BKey_i^{s})$ via BB84
        \State Encrypt $\theta_g^{(t)}$ with $AKey_i^{s}$ to get $Enc(\theta_g^{(t)})$
        \State Send $Enc(\theta_g^{(t)})$ to device $i$
        \State Device $i$ decrypts $Enc(\theta_g^{(t)})$ with $BKey_i^{s}$ to get $\theta_g^{(t)}$
        \State Update $\theta_i^{(t+1)} \gets \theta_g^{(t)}$
    \EndFor
\EndFor
\State \textbf{Output}: Global VQC parameters $\theta_g^{(T)}$
\end{algorithmic}
\end{algorithm}

\subsection{TP QFL}
\begin{figure}
    \centering
    \includegraphics[width=0.7\linewidth]{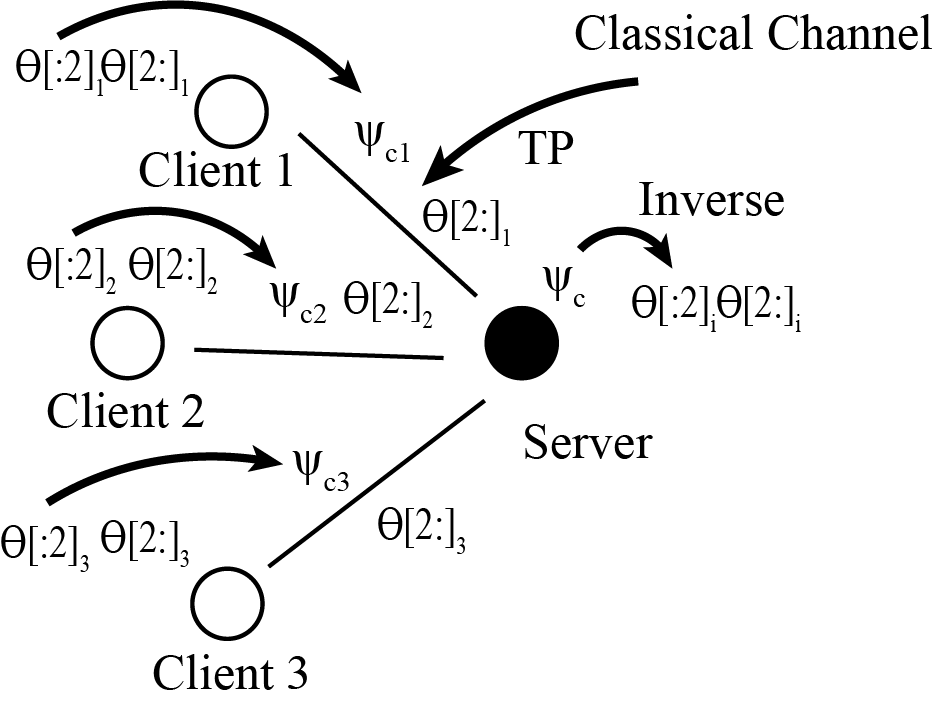}
    \caption{TP QFL}
    \label{fig:tp_qfl}
\end{figure}

We study the aspect of integrating teleportation into a secure QFL framework.
In this approach, a teleportation-based protocol is employed to securely transmit model parameters (weights) from local devices to the a server using quantum teleportation.
In Algorithm \ref{alg:teleportation}, also depicted in Figure \ref{fig:tp_qfl}, in each communication round, each devices $i$ train local VQCs to obtain parameters $\theta_i$.
For each client, we propose the use of the first two parameters $\theta_i[0]$ and $\theta_i[1]$ (variations can be implemented; such as selecting parameter values at different indexes for each device), to be used or encoded as
angles $\theta$ and $\varphi$ in a quantum state in the qubit $Q$.
Each device entangles an auxiliary qubit $A$ with a server-held qubit $B$ using a Hadamard gate and CNOT gate, creating a bell state.
The quantum state $Q$ is then entangled with $A$, followed by measurements on $Q$ and $A$.
The measurement results are classically sent to the server \cite{bennettQuantumCryptographyPublic2014}.

On the server, based on the results received, conditional gates $X$ or $Z$ are applied on the qubit $B$ to reconstruct the teleported state.
To extract the first two parameters, an inverse unitary $U^\dagger(\theta, \varphi)$ is applied and $B$ is measured to verify teleportation. 
Then with a complete set of parameters from each device, the server aggregates the received parameters
$\theta_i$ by computing the mean $\theta_g = \frac{1}{n} \sum_{i=1}^n \theta_i$, which is distributed back to the devices to update their local VQCs. This process repeats for multiple rounds, ensuring secure and efficient parameter sharing via quantum teleportation.

\begin{algorithm}
\caption{TP-QFL}
\label{alg:teleportation}
\begin{algorithmic}[1]
\State \textbf{Input}: $n$ devices, VQC parameters $\theta_i$
\State Initialize local VQC models with random $\theta_i$
\For{each round $t = 1$ to $T$}
    \For{each device $i = 1$ to $n$}
        \State Train local VQC to update $\theta_i^{(t)}$
        \State Encode $\theta_i^{(t)}[j], \theta_i^{(t)}[j+1]$(or $j$ = random) as angles $\theta, \varphi$ on qubit $Q_i$
        \State Create entangled pair $(A_i, B_i)$ using Hadamard and CNOT gates
        \State Entangle $Q_i$ with $A_i$, apply Hadamard on $Q_i$, measure $Q_i$ and $A_i$
        \State Send measurement outcomes to server
    \EndFor
    \State Server receives measurement outcomes and applies conditional gates $X$ and $Z$ gates on $B_i$ based on received outcomes.
    \State Server applies inverse unitary $U^\dagger(\theta, \varphi)$ on $B_i$ and  measure $B_i$ to obtain $\theta_i^{(t)}$
    \State Compute global $\theta_g^{(t)} = \frac{1}{n} \sum_{i=1}^n \theta_i^{(t)}$
    \State Distribute $\theta_g^{(t)}$ to devices
    \For{each device $i = 1$ to $n$}
        \State Update local VQC parameters $\theta_i^{(t+1)} \gets \theta_g^{(t)}$
    \EndFor
\EndFor
\State \textbf{Output}: Global VQC parameters $\theta_g^{(T)}$
\end{algorithmic}
\end{algorithm}

\subsection{KEM QFL}
\begin{figure}
    \centering
    \includegraphics[width=0.9\linewidth]{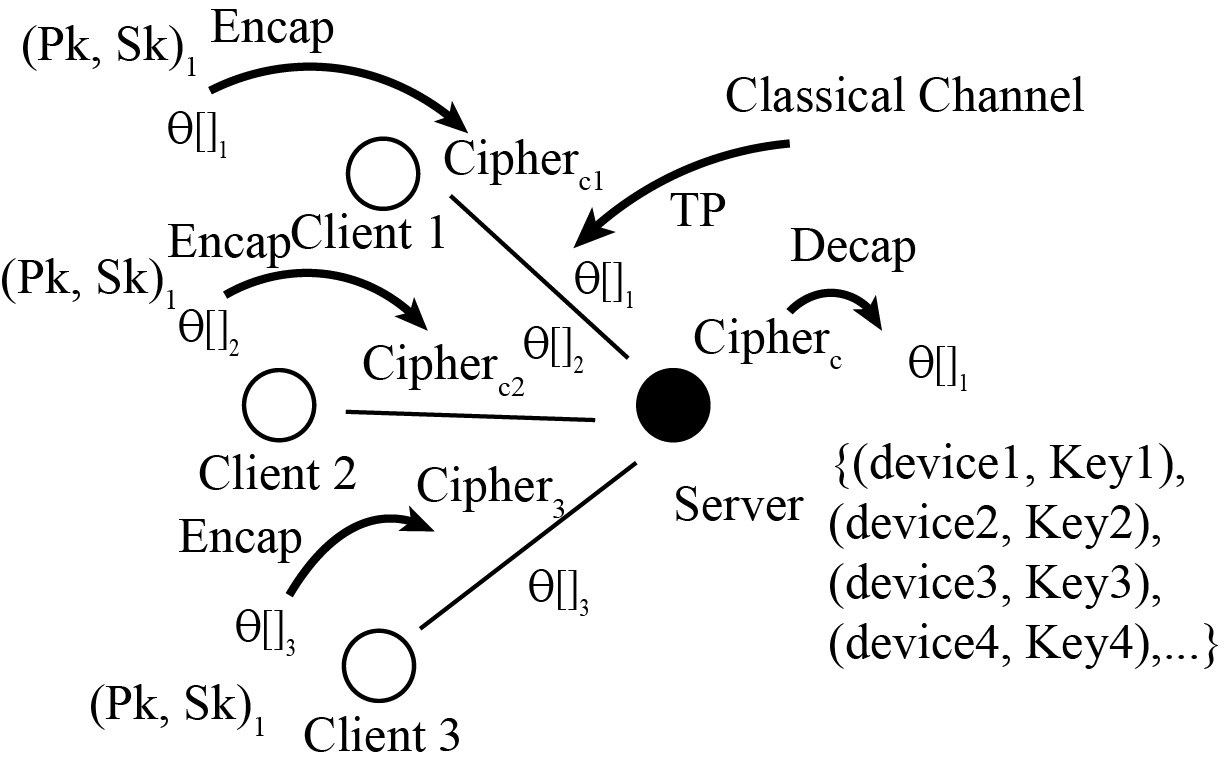}
    \caption{KEM QFL}
    \label{fig:kem_qfl}
\end{figure}

In this work, we present the idea of using postquantum key encapsulation mechanisms (KEMs) such as Kyber to securely share model parameters between devices and a server.
In this approach, as shown in Algorithm \ref{alg:kem}, also shown in Figure \ref{fig:kem_qfl}, similar to other standard QFL procedures, in each round of communication, the $n$ devices train the local VQC model to obtain parameters $\theta_i$.
Each device converts $\theta_i$ to bytes, computes a hash $h(\theta_i)$, and uses the KEM public key of the server $pk_s$ to encapsulate a shared secret $ss_i$ and ciphertext $ct_i$.
The hash $h(\theta_i)$ is encrypted with $ss_i$ using schemes such as AES-GCM, producing a symmetric ciphertext $sct_i$ and nonce. 
The tuple $(ct_i, sct_i, nonce_i)$ is sent to the server. 
The server decapsulates $ct_i$ using its private key $sk_s$ to recover $ss_i$, 
then decrypts $sct_i$ to obtain $h(\theta_i)$. 
The server aggregates the parameters $\theta_g = \frac{1}{n} \sum_{i=1}^n \theta_i$ and distributes $\theta_g$ to devices to update the local VQCs.

\begin{algorithm}
\caption{KEM QFL}
\label{alg:kem}
\begin{algorithmic}[1]
\State \textbf{Input}: $n$ devices, VQC parameters $\theta_i$
\State Initialize local VQC models with random $\theta_i$
\State Server generates KEM key pair $(pk_s, sk_s)$
\For{each round $t = 1$ to $T$}
    \For{each device $i = 1$ to $n$}
        \State Train local VQC to update $\theta_i^{(t)}$
        \State Convert $\theta_i^{(t)}$ to bytes and compute hash $h(\theta_i^{(t)})$
        \State Encapsulate shared secret $ss_i$ and ciphertext $ct_i$ using $pk_s$
        \State Encrypt $h(\theta_i^{(t)})$ with $ss_i$ and nonce $nonce_i$ using AES-GCM to get $sct_i$
        \State Send $(ct_i, sct_i, nonce_i)$ to server
    \EndFor
    \State Server decapsulates each $ct_i$ with $sk_s$ to recover $ss_i$
    \State Decrypt each $sct_i$ with $ss_i$ and $nonce_i$ to recover $h(\theta_i^{(t)})$
    \State Recover $\theta_i^{(t)}$ from $h(\theta_i^{(t)})$ and aggregate $\theta_g^{(t)} = \frac{1}{n} \sum_{i=1}^n \theta_i^{(t)}$
    \For{each device $i = 1$ to $n$}
        \State Distribute $\theta_g^{(t)}$ to device $i$
        \State Update local VQC parameters $\theta_i^{(t+1)} \gets \theta_g^{(t)}$
    \EndFor
\EndFor
\State \textbf{Output}: Global VQC parameters $\theta_g^{(T)}$
\end{algorithmic}
\end{algorithm}

\subsection{PQC QFL}
With post quantum cryptography, we conduct study of their implementation into QFL with various PQC schemes.
PQC schemes such as Dilithium, Falcon1024, ML-DSA024, SPhincs, Mayo are used for public-key cryptography, whereas for key encapsulation mechanism, we perform study with BIKE, McElience, Kyber, KEM1024 and ProdoKEM.
With public key cryptography, 
after completion of the training, each client signs the model parameters and sends the model parameters, signed signature, and the public key is already publicly available.
The server receives this information and performs the verification of the signature to identify if any compromise has been made on the model parameters.

\begin{figure}
    \centering
    \includegraphics[width=0.9\linewidth]{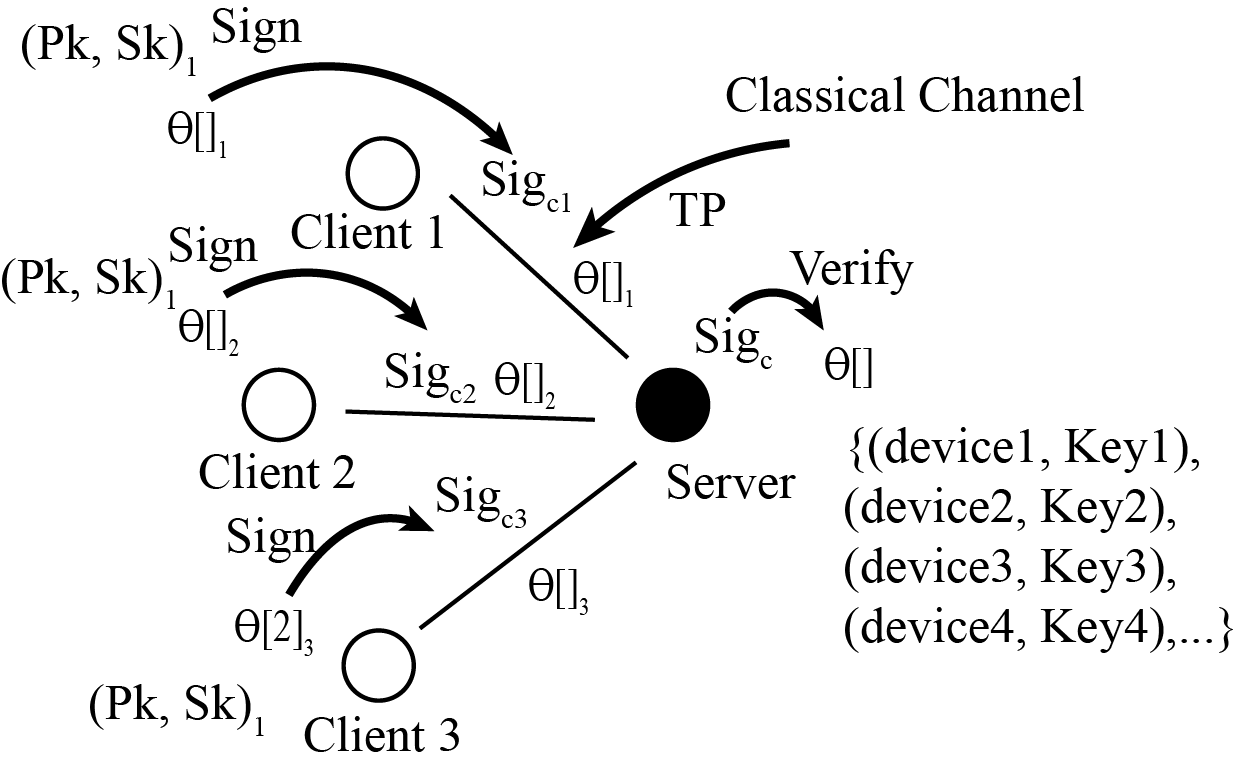}
    \caption{PQC QFL}
    \label{fig:pqc_qfl}
\end{figure}

In Algorithm \ref{alg:pqc}, also depicted in Figure \ref{fig:pqc_qfl},
 $n$ devices train the local VQC model to obtain parameters $\theta_i$. 
 Each device converts $\theta_i$ to bytes and signs it with its private key $sk_i$ using a PQC signature scheme, 
 producing a signature $\sigma_i$. 
 The tuple $(\theta_i, \sigma_i, pk_i)$ is sent to the server, where $pk_i$ is the public key of the device. 
 The server verifies each $\sigma_i$ using $\theta_i$ and $pk_i$ to ensure authenticity. 
 If all signatures are valid, the server aggregates the parameters by computing $\theta_g = \frac{1}{n} \sum_{i=1}^n \theta_i$ and distributes $\theta_g$ to the devices to update the local VQCs.

\begin{algorithm}
\caption{PQC-Secured QFL}
\label{alg:pqc}
\begin{algorithmic}[1]
\State \textbf{Input}: $n$ Devices, VQC parameters
\State Initialize local models on each device with VQC parameters.
\State Each device generates a PQC signature key pair $(pk_i, sk_i)$.
\For{each communication round}
    \For{each device $i \in \{1, \dots, n\}$}
        \State Train local model to obtain parameters $\theta_i$.
        \State Convert $\theta_i$ to bytes and sign with $sk_i$ to obtain signature $\sigma_i$.
        \State Send $(\theta_i, \sigma_i, pk_i)$ to server.
    \EndFor
    \State Server verifies each $\sigma_i$ using $\theta_i$ and $pk_i$.
    \If{all signatures are valid}
        \State Compute global parameters $\theta_g = \frac{1}{n} \sum_{i=1}^n \theta_i$.
        \State Distribute $\theta_g$ to devices for local VQC updates.
    \EndIf
\EndFor
\State \textbf{Output}: Globally aggregated VQC parameters $\theta_g$.
\end{algorithmic}
\end{algorithm}

\section{Theoretical Analysis}
\subsection{QKD}
QKD-secured QFL framework relies on protocols such as BB84 \cite{bennettQuantumCryptographyPublic2014} for the generation of secure keys.
QKD secret key can then be used with symmetric encryption schemes such as Fernet or OTP encryption mechanisms.
This approach improves the security measures for the QFL framework by using quantum principles for key distribution.
Due to the law of quantum physics, such as the no-cloning theorem and Heisenberg's uncertainty principles, 
the sharing of secret key cannot be intercepted since any disturbance in quantum state induces detectable errors.
In addition, BB84 protocols are resistant to quantum algorithms \cite{notavailableSecurityQuantumKey2002}.
For best security, a fresh key pair can be generated per device in each communication round or after each use.
In terms of encryption and decryption, 
Fernet is computationally secure, however, not fully secure against quantum threats.
However, OTP offers the best security to work with QKD secret keys.

OTP cryptosystem is a classical cryptography system that provides security against any adversary no matter how computationally powerful the adversary is given three requirements: keys are truly random, known only to legitimate parties, as long as the message is not reused \cite{williamsExplorationsQuantumComputing2011}.

With QKD, we establish a secure key between two parties through the quantum channel \cite{metgerSecurityQuantumKey2023}. 
Followed by OTP or Fernet encryption, the protocol perfectly works to secure model parameters and communication between server and client.
QKD using a protocol like BB84 \cite{bennettQuantumCryptographyPublic2014} is used to establish a shared secret key $K$ between client and server.
The server requires a collection of private keys for each client.
If we have $\mathcal{H}$ as a Herbert space of quantum states such as qubit states $|0\rangle, |1\rangle, |+\rangle, |-\rangle$, then
each client sends a sequence of $n$ qubits, each prepared in a random state $|\psi_i \rangle \in \{|0\rangle, |1\rangle, |+\rangle, |-\rangle\}$, where
$ |+\rangle = \frac{|0\rangle + |1\rangle}{\sqrt{2}} $, $ |-\rangle = \frac{|0\rangle - |1\rangle}{\sqrt{2}}$.
After receiving, the server measures each qubit received from the clients by choosing a basis (rectilinear $\{|0\rangle, |1\rangle\}$ or diagonal $ \{|+\rangle, |-\rangle\} $).
Then, the client and the server communicate and decide which bases were used and only keep the ones that match that produces a key.

Once the private key is successfully shared, 
the key is then used with the OTP or Fernet protocol to encrypt the model parameters.
Let us suppose $M$ as the model parameters in byte form or as $M \in \{0,1\}^l$
with $l$ as the length of the string.
With BB84 like protocol, the key generation involves
random string generation where a quantum circuit with \( n \) qubits is initialized. 
Each qubit is put into a superposition state using the Hadamard gate
$H|0\rangle = \frac{|0\rangle + |1\rangle}{\sqrt{2}}.$
Measurement on a computational basis \( \{|0\rangle, |1\rangle\} \) yields a random bit string \( K \in \{0,1\}^n \), with
$P(0) = P(1) = \frac{1}{2}$.
The client and the server generate random rotation strings \( R_A, R_B \in \{0,1\}^n \). 
For each qubit \( i \)
\[
|\psi_i\rangle = H^{R_B[i]} H^{R_A[i]} X^{K[i]} |0\rangle,
\]
where, each bit $ K[i] $, $ R_A[i] $, $ R_B[i] $ is either 0 or 1, \( X^0 = I \), \( H^0 = I \), and \( X \) is the Pauli-X gate. 
The measurement yields Bob's result string as \( B \in \{0,1\}^n \).
For positions \( i \) where \( R_A[i] = R_B[i] \), the corresponding bit is kept as
\[
K_{\text{final}} = \{ K[i] \mid R_A[i] = R_B[i] \}.
\]
The final key \( K_{\text{final}} \in \{0,1\}^k \), with \( k \approx n/2 \) (approximately half of the qubits contribute to the final key), is derived.

With OTP, the encryption process begins with the preparation of the message, where the model weights $W \in \mathbb{R}^d$ are rounded to certain decimal places $dp$ and serialized into a proper string format such as JSON
as $M = \text{jsonDumps}(\text{round}(W, dp))$ (in the experiment), with $M$ being a string of length $l$. 
Next, in the encryption phase, the key is truncated to $K_A = A_{\text{key}}[:l]$. 
For each character $m_i \in M$ and the corresponding key character $k_i \in K_A$, 
encryption is performed as $c_i = \text{chr}((\text{ord}(m_i) + 2 \cdot \text{ord}(k_i)) \mod 256)$, 
producing the ciphertext $C = c_1 c_2 \dots c_l$. 
During decryption, using the key $K_B = B_{\text{key}}[:l]$, 
each ciphertext character $c_i \in C$ is decrypted via $m_i = \text{chr}((\text{ord}(c_i) - 2 \cdot \text{ord}(k_i)) \mod 256)$. 
This recovers the original string $M = \{m_1 m_2 \dots m_l\}$, 
which is then parsed back into the model weights using $W = \text{jsonLoads}(M)$.

With Fernet, the encryption begins with message preparation as before.
In the encryption phase, using a 256-bit key $K_A$, 
a Fernet object is initialized as $f = \text{Fernet}(K_A)$, 
and the message is encrypted as $C = f.\text{encrypt}(M.\text{encode}())$.
During decryption, 
using the key $K_B$, a Fernet object is initialized
as $f = \text{Fernet}(K_B)$, and the message is
recovered as $M = f.\text{decrypt}(C).\text{decode}()$. 
The model weights are then interpreted as $W = \text{jsonLoads}(M)$.

To integrate encryption into 
QFL, 
the process begins with key generation, 
where devices and the server generate key pairs $(A_{\text{key}}, B_{\text{key}})$ 
via QKD, with the key length $n$ defined.
In the weight processing phase, the devices encrypt their weights $W_i$ using $A_{\text{key}}$, 
and the server decrypts these weights using $B_{\text{key}}$. 
The server then computes the average weights as $W_{\text{avg}} = \frac{1}{N} \sum_{i=1}^N W_i$. 
Subsequently, the server encrypts $W_{\text{avg}}$ and sends it
to the devices, which decrypt it using $B_{\text{key}}$.

\subsection{Teleportation}
The quantum teleportation protocol transfers an unknown quantum state from a sender (Alice) to a receiver (Bob) (QFL requires multiple concurrent server client pair protocols) using an entangled pair of qubits and classical communication. This is implemented with a quantum circuit involving three qubits: a secret qubit (\( Q \)), Alice’s qubit (\( A \)), Bob’s qubit (\( B \)), and a classical register (\( \text{cr} \)) with three bits.
Let the secret state to be teleported be
\[
|\psi\rangle_Q = \alpha |0\rangle + \beta |1\rangle,
\]
where \( \alpha, \beta \in \mathbb{C} \), and \( |\alpha|^2 + |\beta|^2 = 1 \). 
The protocol proceeds by
entangling Alice and Bob
by creating an entangled Bell state between Alice’s qubit \( A \) and Bob’s qubit \( B \) as \cite{williamsExplorationsQuantumComputing2011}
\[
|\Phi^+\rangle_{AB} = \frac{|0\rangle_A |0\rangle_B + |1\rangle_A |1\rangle_B}{\sqrt{2}}
\]
which is achieved by applying a Hadamard gate to \( A \) as
\[
H|0\rangle_A = \frac{|0\rangle_A + |1\rangle_A}{\sqrt{2}},
\]
followed by a CNOT gate with \( A \) as the control and \( B \) as the target.
\[
\text{CNOT}_{A \rightarrow B} \left( \frac{|0\rangle_A + |1\rangle_A}{\sqrt{2}} \otimes |0\rangle_B \right) = \frac{|0\rangle_A |0\rangle_B + |1\rangle_A |1\rangle_B}{\sqrt{2}}.
\]
The initial state is thus 
\[
|\psi\rangle_Q \otimes |\Phi^+\rangle_{AB} = (\alpha |0\rangle + \beta |1\rangle)_Q \otimes \frac{|0\rangle_A |0\rangle_B + |1\rangle_A |1\rangle_B}{\sqrt{2}}.
\]
The secret qubit \( Q \) is prepared in state \( |\psi\rangle_Q = \alpha |0\rangle + \beta |1\rangle \) using a unitary rotation as
\[
U(\theta, \varphi, 0) |0\rangle = \cos\left(\frac{\theta}{2}\right) |0\rangle + e^{i\varphi} \sin\left(\frac{\theta}{2}\right) |1\rangle,
\]
where \( \theta, \varphi \) are parameters defining \( \alpha = \cos(\theta/2) \), \( \beta = e^{i\varphi} \sin(\theta/2) \).
Then, we
apply a CNOT gate with \( Q \) as control and \( A \) as target, followed by a Hadamard gate on \( Q \)
\[
\text{CNOT}_{Q \rightarrow A} \left( (\alpha |0\rangle + \beta |1\rangle)_Q \otimes |\Phi^+\rangle_{AB} \right),
\]
\[
H_Q \left( \text{CNOT}_{Q \rightarrow A} \\
\left( (\alpha |0\rangle + \beta |1\rangle)_Q \otimes |\Phi^+\rangle_{AB} \right) \right).
\]
The resulting state is
\begin{align*}
|\Psi\rangle = \frac{1}{2} \Big( &|00\rangle_{QA} (\alpha |0\rangle + \beta |1\rangle)_B \\
&+ |01\rangle_{QA} (\alpha |1\rangle + \beta |0\rangle)_B \\
&+ |10\rangle_{QA} (\alpha |0\rangle - \beta |1\rangle)_B \\
&+ |11\rangle_{QA} (\alpha |1\rangle - \beta |0\rangle)_B \Big).
\end{align*}
Now, we
measure qubits \( Q \) and \( A \) in the computational basis, storing the results in classical bits \( \text{cr}[0] \) and \( \text{cr}[1] \) as
\[
\text{Measure } Q \rightarrow \text{cr}[0], \quad \text{Measure } A \rightarrow \text{cr}[1].
\]
The measurement collapses the state to one of four possibilities, e.g., for outcome \( \text{cr}[0] = m_1 \), \( \text{cr}[1] = m_2 \) as
$|m_1 m_2\rangle_{QA} \otimes |\psi_{m_1 m_2}\rangle_B,$
where \( |\psi_{m_1 m_2}\rangle_B \) is one of the following.
\[
\begin{cases}
    \alpha |0\rangle + \beta |1\rangle & \text{if } (m_1, m_2) = (0,0), \\
    \alpha |1\rangle + \beta |0\rangle & \text{if } (m_1, m_2) = (0,1), \\
    \alpha |0\rangle - \beta |1\rangle & \text{if } (m_1, m_2) = (1,0), \\
    \alpha |1\rangle - \beta |0\rangle & \text{if } (m_1, m_2) = (1,1).
\end{cases}
\]
Then, we apply corrections to Bob’s qubit \( B \) based on the measurement results as
\[
\text{If } \text{cr}[1] = 1, \text{ apply } X_B; \quad \text{If } \text{cr}[0] = 1, \text{ apply } Z_B.
\]
The operators correct Bob’s state as
\[
\begin{cases}
    I_B & \text{if } (m_1, m_2) = (0,0), \\
    X_B & \text{if } (m_1, m_2) = (0,1), \\
    Z_B & \text{if } (m_1, m_2) = (1,0), \\
    Z_B X_B & \text{if } (m_1, m_2) = (1,1).
\end{cases}
\]
After correction, Bob’s state becomes
\[
|\psi\rangle_B = \alpha |0\rangle + \beta |1\rangle.
\]
Then, we measure Bob’s qubit \( B \) in the computational basis, storing the result in \( \text{cr}[2] \) as
\[
\text{Measure } B \rightarrow \text{cr}[2].
\]
This verifies the teleported state, but is not strictly necessary for teleportation.
To retrieve the value of the first and second model parameters, we
can apply the inverse of the initial rotation to Bob’s qubit
\[
U(\theta, \varphi, 0)^{-1}_B.
\]

The protocol assumes ideal quantum channels and no eavesdropping. The entangled state \( |\Phi^+\rangle_{AB} \) ensures perfect correlation, and the classical communication of \( \text{cr}[0], \text{cr}[1] \) allows Bob to reconstruct \( |\psi\rangle \). Security is based on the no-cloning theorem, which prevents unauthorized copying of \( |\psi\rangle_Q \).

\subsection{Post-Quantum Cryptography (PQC)}
With postquantum cryptography (PQC) schemes, key generation involves generating a pair of keys $(PK, SK) \leftarrow \text{Gen}_{\text{sig}}()$, where $PK$ is the public key, 
$SK$ is the secret key and $\text{Gen}_{\text{sig}}$ is the key generation algorithm
for the chosen signature scheme. 
For signing,
the process involves serializing
model weights $W \in \mathbb{R}^d$ to a byte string.
The signature is then computed as $\sigma \leftarrow \text{Sign}_{\text{sig}}(SK, M)$, 
where $\text{Sign}_{\text{sig}}$ is the signing algorithm, such as Dilithium’s lattice-based signing. 
Verification 
involves the server checking the signature via $b \leftarrow \text{Verify}_{\text{sig}}(PK, M, \sigma)$, 
where $b \in \{\text{True}, \text{False}\}$ indicates whether the signature is valid. 
This verification ensures the integrity and authenticity of $M$, 
confirming that $W$ originated from the device with the public key $PK$. 
The signature scheme is post-quantum secure, resistant to attacks by quantum computers, such as Shor's algorithm, assuming that the underlying hard problem, such as lattice-based problems, remains intractable.
In scenarios like QFL, we require multiple pairs of server-client key pairs.

\subsection{KEM}
The implemented KEM establishes a shared secret between devices and the server, which is used for the symmetric encryption of the hashed model weights. The KEM scheme (e.g., Kyber) is post-quantum secure.
Key Generation involves, a KEM key pair generation as, 
$(PK_{\text{KEM}}, SK_{\text{KEM}}) \leftarrow \text{Gen}_{\text{KEM}}()$,
where, \( PK_{\text{KEM}} \) and \( SK_{\text{KEM}} \) are the public and secret keys, respectively.

Encapsulation 
involves the device that encapsulates a shared secret using the public key of the server \( PK_{\text{KEM, server}} \): $(C, SS_R) \leftarrow \text{Encap}_{\text{KEM}}(PK_{\text{KEM, server}})$, where \( C \) is the ciphertext and \( SS_R \) is the shared secret (receiver copy).
Decapsulation
involves
the server decapsulating the ciphertext using its secret key as
$SS_S \leftarrow \text{Decap}_{\text{KEM}}(C, SK_{\text{KEM, server}})$,
where \( SS_S \) is the server’s copy of the shared secret.
If successful, \( SS_S = SS_R \), ensuring a shared secret between the device and the server.

Symmetric Encryption involves the following steps.
The first message preparation is performed for the model weights \( W \in \mathbb{R}^d \) to be converted to bytes as 
    \[
    M_{\text{bytes}} = \bigoplus_{i=1}^d \text{pack}(w_i),
    \]
    where,  \( \text{pack}(w_i) \) converts each float \( w_i \) to 8 bytes (double precision, big-endian).
Then, the SHA-256 hash of the weights is computed as 
    \[
    H = \text{SHA256}(M_{\text{bytes}}).
    \]


Then we derive a 256-bit symmetric key $K_{sym}$ using tools like HKDF with SHA-256.
The encryption part includes encrypting the hash \( H \) using AES-GCM with a 12-byte nonce
as 
\[
C_{\text{sym}} = \text{AES-GCM}_{\text{Enc}}(K_{\text{sym}}^{client}, H, \text{nonce}).
\]

Now for decryption, we
derive the symmetric key using the shared server secret.
The ciphertext is encrypted as
\[
H = \text{AES-GCM}_{\text{Dec}}(K_{\text{sym}}^{server}, C_{\text{sym}}, \text{nonce}).
\]
Finally, the server verifies the integrity of \( H \) against a locally computed hash of the received weights.

\section{Security Analysis}

\subsection{QKD}


QKD allows two parties to exchange provably secure keys through a potential insecure quantum channel \cite{korzhProvablySecurePractical2015}.
We extend unconditional security proof of the BB84 QKD protocol against any general attack allowed by quantum mechanics, assuming eavesdropper (Eve) with unlimited computational power and quantum capabilities as presented by Biham et al. \cite{bihamProofSecurityQuantum2000}.
From \cite{bihamProofSecurityQuantum2000}, assumptions for the security proof of QKD include:
\begin{assumption}[Unjammable Classical Channel]
    Sender and receiver share an unjammable classical channel (or authenticated using a short secret key used to authenticate a standard classical channel).
\end{assumption}
\begin{assumption}[Eve Attack]
EveDropper attacks only the quantum channel and listens to all the transmission on classical channel but cannot attack labs, i.e. labs are secure.
\end{assumption}
\begin{assumption}[Two Level System]
The sender sends quantum qubits.
\end{assumption}
\begin{assumption}
    The Eve has unlimited technology (quantum memory, computer).
\end{assumption}

In the BB84 protocol, both sender and receiver use 4 possible quantum states (BB84 states) in 3 bases using the ``spin" notation and connecting them to the ``computational basis" notation, such as  $|0_z\rangle = |0\rangle$
,$|1_z\rangle = |1\rangle$
,$|0_x\rangle = \frac{1}{\sqrt{2}}(|0\rangle + |1\rangle)$
$|1_x\rangle = \frac{1}{\sqrt{2}}(|0\rangle - |1\rangle)$.
In this case, after receiving the states and comparing bases after the sender transmits these states, a common key is created in instances where the sender and the receiver use the same basis.
This is followed by creating a shifted key and creating a final key from the shifted key.

Now, to model the threat, the possible ways Eve can attack qubits is that first it can perform a unitary transformation on sender's qubit and its own probe.
This probe is kept in memory and is used only after all classical information is received from both sender and receiver such as bases of all bits, choices of test bits, test bit values etc.
Using this information, an optimal measurement on the probe is performed to infer as much as possible about the final secret key.
We can now write for $Q_i$ quantum systems that sender sends to the receiver, Eve could collect all of the $n$ systems performing arbitrary quantum channel $\mathcal{A}: Q^n \rightarrow EQ^n$ (General Attack) and send output on systems $Q^n$ ($n$ quantum systems) to the receiver. \cite{metgerSecurityQuantumKey2023}.
\begin{assumption}
    Eve can possess only one quantum system $Q_i$ at a time.
\end{assumption}
The sender sends the systems $ Q_1, Q_2, ..., Q_n $ sequentially, which forces Eve's attack to also be sequential. 
Thus, the attack is modeled as a sequence of maps $ A_i: E'_{i-1} Q_i \to E'_i Q_i $, where, 
$ E'_{i-1} $ is Eve’s side information before intercepting $ Q_i $, 
$ E'_i $ is her updated side information after processing $ Q_i $ and
$ Q_i $ is sent to Bob after Eve’s operation.

For QFL setting with $N$ clients, $T$ communication rounds, the full attack can be given by the tensor product map, 
\[
\mathcal{A}^{\otimes n}: Q^n \rightarrow EQ^n  \quad \forall N, T
\]

The security and reliability of QKD can be further analyzed as follows.
Given a test is passed, Eve's information about Alice's key (sender) key is exponentially small as, 
\[
I(A; E | T = \text{pass}) < A_{\text{info}} e^{-\beta_{\text{info}} n}
\]
where, $I(A; E | T = \text{pass})$
is the mutual information between Alice's key (A) and Eve's information (E), conditioned on test passing, $n$ is the number of qubits used in the protocol.
But due to attacks such as SWAP attacks, this criterion fails.
In brief, during swap attack, Eve intercepts Alice's qubits and stores them in quantum memory (without measuring them immediately).
Then Alice sends random BB84 states (qubits in random bases and polarization) to Bob.
After Alice and Bob publicly announce the bases used during the test phase, Eve measures the stored qubits in the correct bases and gains the full information about Alice's key.
But the problem with this is that, when the test passes (a rare event), Eve's information is not small and is equal to the key length (m bits).
Thus, the above bound cannot be satisfied, as Eve's information does not decrease exponentially.
Similarly, the half-SWAP attack also fails where Eve, with a probability of $\frac{1}{2}$, does nothing to Alice's qubits to reach Bob undisturbed (test passes with high probability at Bob).
With another half probability, Eve performs SWAP attack, i.e. intercepts Alice's qubits, sends random states to Bob, and measures Alice's qubits later.
So, on average, $I(A;\varepsilon) = \frac{m}{2}$ and
$P(T=pass) \geq \frac{1}{2}$.

Thus, the correct security criterion should control the joint probability that both bad things happen at the same time i.e. Eve gains non-negligible information ($ I_{\text{Eve}} \geq A_{\text{info}} e^{-\beta_{\text{info}} n} $) and protocol passes the test (T=pass).
Thus, the security criterion is \cite{bihamProofSecurityQuantum2000}:
\[
P \left[ (T = \text{pass}) \land (I_{\text{Eve}} \geq A_{\text{info}} e^{-\beta_{\text{info}} n}) \right] < A_{\text{luck}} e^{-\beta_{\text{luck}} n}
\]
where, let \( T \) be the result of the test and \( I_{\text{Eve}} = I(A; E \mid i_T, C_T, b, s) \) be the information Eve has about the key, with the protocol parameters \( (i_T, j_T, b, s) \) disclosed by Alice and Bob. 
The test is considered passed when \( C_T = i_T \sim j_T \) satisfies \( |C_T| \leq n_{\text{allowed}} \). 
Alice and Bob can increase the number of bits \( n \) to increase security.
This means that the probability that the protocol passes the test and Eve has significant information is exponentially small.
Also, with respect to reliability, the probability that each client and server's final keys differ is exponentially small as 
\[
P(A \neq B) \leq A_{rel}e^{-B_{rel}n}
\]

The most general attack by Eve can be such that Eve attaches a quantum probe and performs a unitary transformation $U$ on all the qubits and her probe and then sends disturbed qubits to receiver.
Later, when the sender and receiver publish all classical data, she measures her probe optimally.
The unitary transformation $U$ is written as follows. 
\[
U(|0\rangle|i\rangle) = \sum_j |E'_{i,j}\rangle|j\rangle
\]
where, $|E'_{i,j}\rangle$ are the unnormalized states of Eve's probes when sender sends $|i\rangle$ and the receiver receives $|j\rangle$, $i$ is a string encoded in the bases of her choice $b$ by sender and the receiver measures a string $j$ using the same set of bases, $|0\rangle$ is probe in unknown state prepared by Eve.

For QKD to be secure, it satisfies the following:
\begin{definition}[Correctness, Secrecy and Completeness]
    \begin{enumerate}
        \item Correctness. Sender's and Receiver's final keys $K$ and $K'$ are equal except with probability $\leq \varepsilon_{cor}$ given the protocol doesn't abort ensuring robustness against adversarial interference
        \[
        Pr[K \neq K' | \text{not abort}] \leq \varepsilon_{cor}
        \]

        \item Secrecy. Eve's information about the final key is negligible and thus the trace norm difference between the actual state $\rho_{KE|\Omega}$ (conditioned on not aborting) and an ideal state where the key is maximally mixed $\tau_{K\otimes\rho E}$ is bounded as, 
        \[
        \|\rho_{KE|\Omega} - \tau_K \otimes \rho_{E|\Omega}|\\| \leq \epsilon_{\text{sec}}.
        \]

        $\tau_K$ is the maximally mixed state on system $K$, $\Omega$ is the event that the protocol doesn't abort.

        \item Completeness. There exists an honest behavior for the adversary Eve for a given noise model for the protocol such that, 
        \[
        Pr[\text{abort}] \leq \varepsilon^{comp}.
        \]
        This ensures robustness against given noise model and the probability of aborting the protocol is small if Eve behaves honestly.

    \end{enumerate} 
\end{definition}

Thus, the QKD protocol is $(\varepsilon_{\text{cor}} + \varepsilon_{\text{sec}}/2)$-secure.

\subsection{Teleportation}
For distributed systems like QFL, quantum teleportation between remote and far distant nodes in a quantum network requires remote entanglement between multiple pairs of server client nodes, entanglement swapping, storage in memory, etc. \cite{hermansQubitTeleportationNonneighbouring}.
One of the key parameters for quantum teleportation is the fidelity of the preshared entangled states between the clients and the servers \cite{hermansQubitTeleportationNonneighbouring}.

Due to the no-cloning theorem, the original state in the sender is destroyed \cite{bouwmeesterExperimentalQuantumTeleportation1997}.
Also, sender's measurement results and the classical information of the receiver do not reveal anything and thus leak no information about the teleported quantum state and their amplitudes $a$ and $b$.
The Bell state measurement performed by the sender projects its two qubits into four maximally (loose all information when considered alone) entangled states as follows: 
\[
|\Psi^{\pm}\rangle, |\Phi^{\pm}\rangle.
\]
This operation destroys the original quantum state , 
\[
|\psi\rangle \rightarrow \text{Destroyed}
\]

The classical message consists of two bits (00, 01, 10 or 11) indicating which Bell state was detected.
These bits do not carry information about $|\psi\rangle$.
Similarly, entangled pair $|\Psi^-\rangle$ is maximally mixed locally i.e. any single qubit individually has completely random density matrix as 
\[
\rho_2 = \rho_3 = \frac{I}{2}
\]
where, $I$ is the identity matrix $I = \begin{pmatrix} 1 & 0 \\ 0 & 1 \end{pmatrix}, \quad \frac{I}{2} = \begin{pmatrix} \frac{1}{2} & 0 \\ 0 & \frac{1}{2} \end{pmatrix}$.
Thus, information theoretically, the protocol is secure against classical eavesdropping.
Even with full access to both the quantum channel and the classical bits, Eve cannot reconstruct or estimate $|\psi\rangle$ without possessing the entangled qubit of the receiver.

Thus, teleportation is secure against various threats, such as
Eavesdropping on the classical channel. 
The classical bits reveal no information about $a$ and $b$, as they are random Bell state labels.
Similarly, even if Eve captures either of the qubits during entanglement distribution, each qubit alone as maximally mixed has no useful information alone.
Attack on both classical and entangled particles is also still insufficient.

\subsection{PQC}
Post quantum cryptography including public key cryptography and key encapsulation mechanisms is based on various mathematical problems such as lattice-based Module Learning with Errors (MLWE) problem, code-based or hash-based etc.
\cite{paarPostQuantumCryptography2024}.
The PQC signature schemes vary according to the mathematical problems that they depend on.
Such as the PQC scheme Crystals Dilithium, its security is based on the hardness of problems defined over the polynomial ring $R_q = \mathbb{Z}[X]/(X^n + 1)$, where $q$ is a prime modulus and $\mathbb{Z}_q$ is the ring of integers modulo $q$. The problems are MLWE , MSIS (Module Short Integer Solution) etc. 

Let us suppose that the advantage of an adversary $A$ breaking the SUF-CMA (Strong Unforgeability under Chosen-Message Attack) security of the signature scheme $scheme$ is $Adv^{SUF-CMA}_{Scheme} (A)$ based on the hardness of problems such as $Problem1, Problem2, ..$ then \cite{ducasCRYSTALSDilithiumLatticeBasedDigital2018},
\[
Adv^{SUF-CMA}_{Scheme} (A) \leq Adv^{Problem}_{params}(P1) + \cdots + 2^{-254}
\]

For $N$ clients, the signature scheme protocol needs to secure the communication with each client-server communication, which will also be $N$ pairs of communications.

\subsection{KEM}
KEM is a type of key establishment scheme where participating parties need to establish a shared key (symmetric key) before symmetric cryptography can be used.
\cite{alagicRecommendationsKeyEncapsulationMechanisms2025}.
It consists of 3 stages: First is key generation for private (decapsulation) key $dk$ and public (encapsulation) key $ek$ on input $1^k$ as, 
\[
(ek, dk) \leftarrow Gen(1^k; r_g)
\]
Now the receiver uses the sender's encapsulation key to generate a shared secret key $K$ and an associated ciphertext $c$ as, 
\[
(c,K) \leftarrow  Encaps(ek;re) 
\]
The ciphertext is then sent to the sender.
Sender then uses ciphertext and decapsulation key to compute the copy of the shared secret key $K'$ as 
\[
K' \leftarrow Decaps(p, dk, c)
\]

\begin{definition}[Correctness]
   KEM protocol = (Gen, Encaps, Decaps) is correct if, for any (ek,dk) generated by Gen, we have  \cite{saitoTightlySecureKeyEncapsulationMechanism2018}
   \[
   Pr[Decaps(dk,c) = K : (c,K) \leftarrow Encaps(ek)] = 1
   \]
\end{definition}

\begin{definition}[Seurity of KEM]
    For any computationally bounded adversary $A$ that receives $(ek, c, K_b)$, KEM is IND-CPA (Indistinguishability under Chosen-Plaintext Attack) secure if the difference between the probability that $A$ wins the experiment IND-CPA and $\frac{1}{2}$ is negligible as \cite{alagicRecommendationsKeyEncapsulationMechanisms2025}:
    \[
    |Pr[A \text{ wins } IND-CPA] - \frac{1}{2}| \leq \text{negl}(n).
    \]
    Adversary $A$ is free to study the encapsulation key $ek$ and ciphertext $c$ in order to identify whether $K_b$ is the true key. 
    Similarly, with IND-CCA (In-distinguishability under Chosen-Ciphertext Attack), adversary now has oracle access to Decaps(p, dk, c) which means adversary is permitted to submit ciphertexts $c^*$ that they generate and get the response $K^* \leftarrow \text{Decaps}(p, dk, c*)$. 
    Then, KEM is IND-CCA if, for every computationally bounded adversary $A$, 
    \[
    |Pr[A \text{ wins } IND-CCA] - \frac{1}{2}| \leq \text{negl}(n).
    \]
\end{definition}


\section{Experimental Results}
We present various experimental analyses with QKD, Teleportation, KEM and PQC schemes with protocols applicable to QFL networks.

\subsection{Set Up}
In this experiment, each client has the Variational Quantum Classifier (VQC), a local optimizer, and its own local dataset.
We have performed experimental analysis on two datasets: IRIS and Genomic.
For data distribution, the whole dataset is divided into test set and training set.
The test set is for the server device for validation and testing, whereas the training set is shared among the clients.
Within each client, the data are further split into train and validation set for training and validation within the local device.
We use normal data size for IRIS and for Genomic, we select 5000 samples for training and 150 samples for server.
The experiment involves 3 client devices with IRIS, whereas with genomic data, we use 20 client devices.
Genomic data are dimensionally reduced to 4 features using Principal Component Analysis.
Within each device, we have the 3 layer ansatz circuit (RealAmplitudes), 1 layer feature map (ZFeatureMap), the COBYLA optimizer with max iteration of 10, and the AerSimulator as noisy quantum circuit simulator backend.
The experimental data distribution is IID in nature, with experiment run for 10 and 20 communication rounds with IRIS and Genomic data, respectively.
1024 sampler shots were used for circuit simulation.
For PQC and KEMs algorithms, we used the liboqs\footnote{\url{https://github.com/open-quantum-safe/liboqs-python}} Python library.
For QKD we follow the BB84 protocol \cite{bennettQuantumCryptographyPublic2014}.
All experimental analysis is based on the Qiskit\footnote{\url{https://github.com/Qiskit/qiskit}} library.

\subsection{Results}
\subsubsection{QKD}

\begin{figure}[!htbp]
    \centering
    \begin{subfigure}[b]{0.32\columnwidth}
        \centering
       \includegraphics[width=\columnwidth]{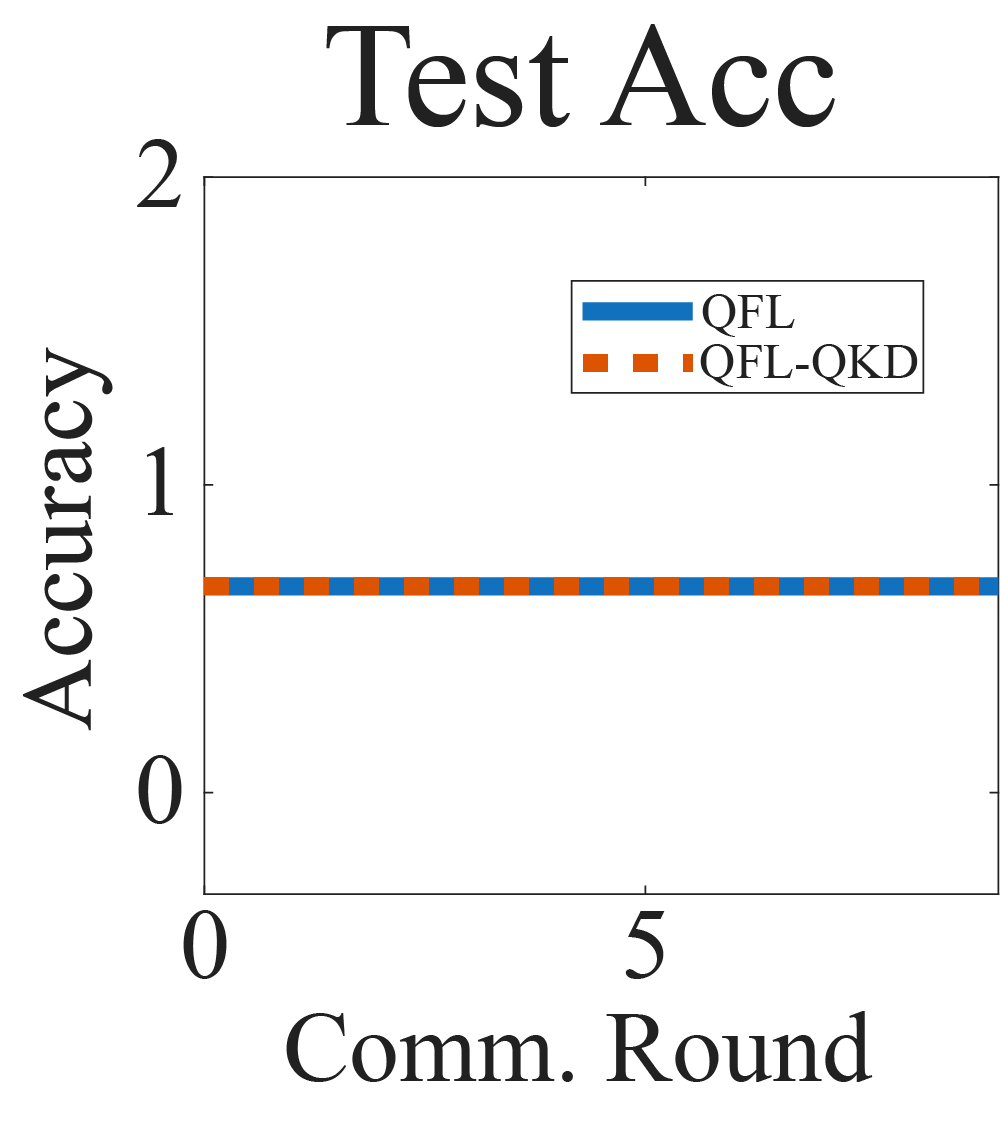}
    \caption{Server Acc}
    \label{fig:server_test_acc_qkd}
    \end{subfigure}
       \begin{subfigure}[b]{0.32\columnwidth}
        \centering
       \includegraphics[width=\columnwidth]{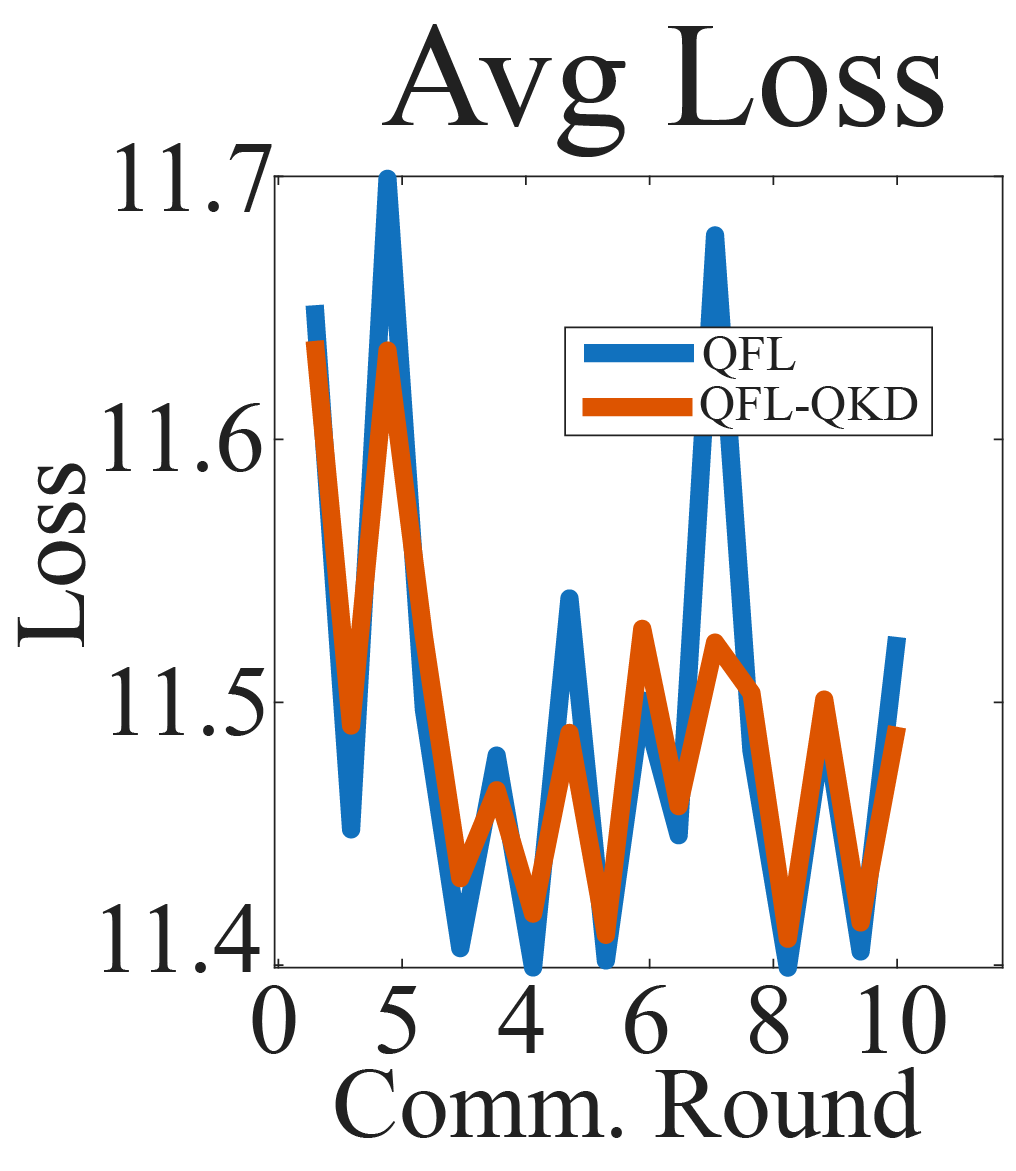}
    \caption{Avg Devices Loss}
    \label{fig:avg_devices_obj_values_qkd}
    \end{subfigure}
    \begin{subfigure}[b]{0.32\columnwidth}
        \centering
       \includegraphics[width=\linewidth]{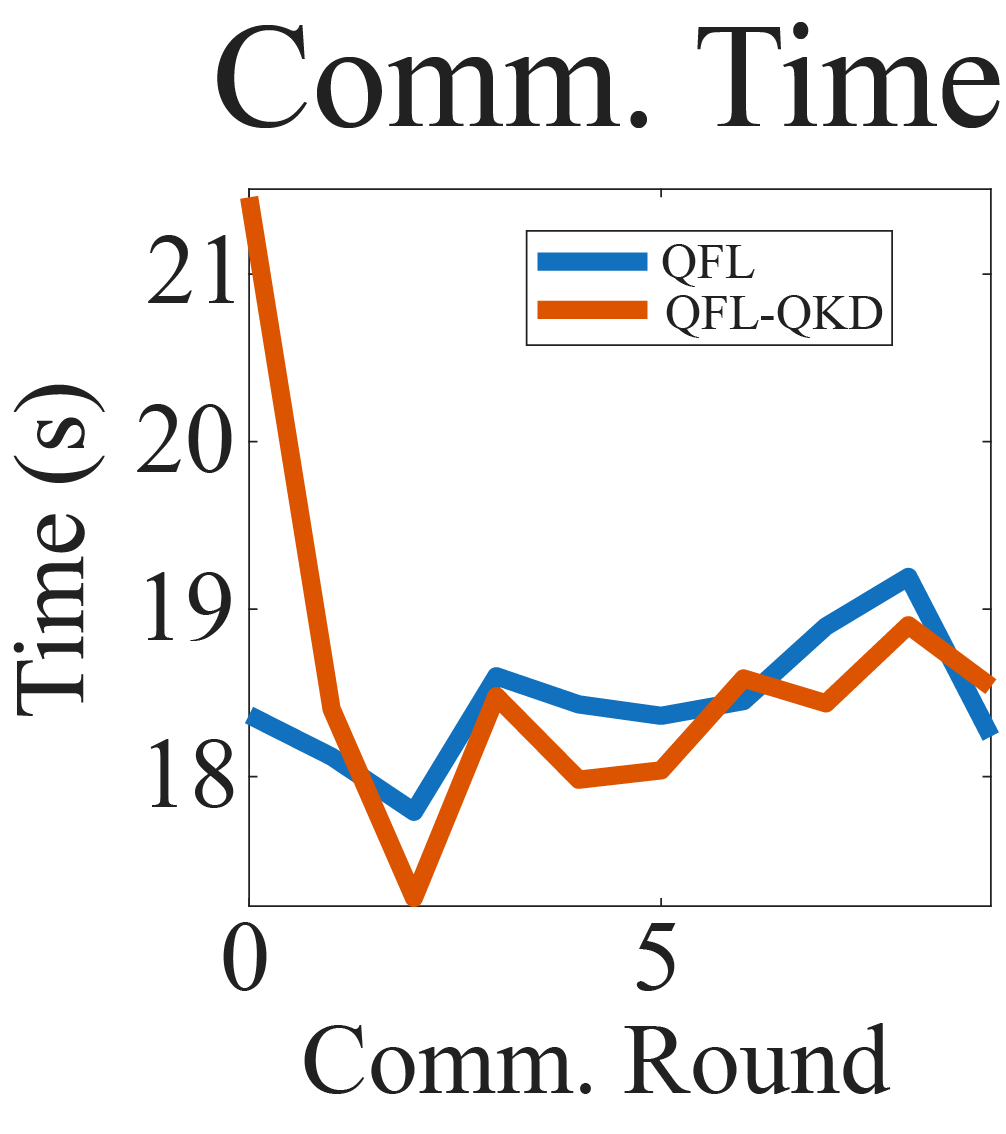}
    \caption{Comm Time} 
    \label{fig:comm_time_qkd}
    \end{subfigure}   
    \caption{Overall Performance: QFL vs QFL-QKD}
    \label{fig:server_performance_qkd}
\end{figure}

Figure \ref{fig:server_performance_qkd} shows the server performance in 10 communication rounds.
Even though QKD implementation is more for security purposes, their is variation in the performance with or without QKD integration.
However, it is hard to confirm direct causality between server performance and QKD integration, since QKD is only used for encryption and decryption of QFL parameters.
These are more observational results rather than comparisons, such as which performs better, since except for QKD integration, everything is the same. 
However, only with integration of QKD, there is security implementation.

\subsubsection{QFL vs QFL QKD vs QFL QKD Fernet vs QFL TP}
\begin{figure}[!htbp]
    \centering
    \begin{subfigure}[b]{0.32\columnwidth}
        \centering
       \includegraphics[width=\columnwidth]{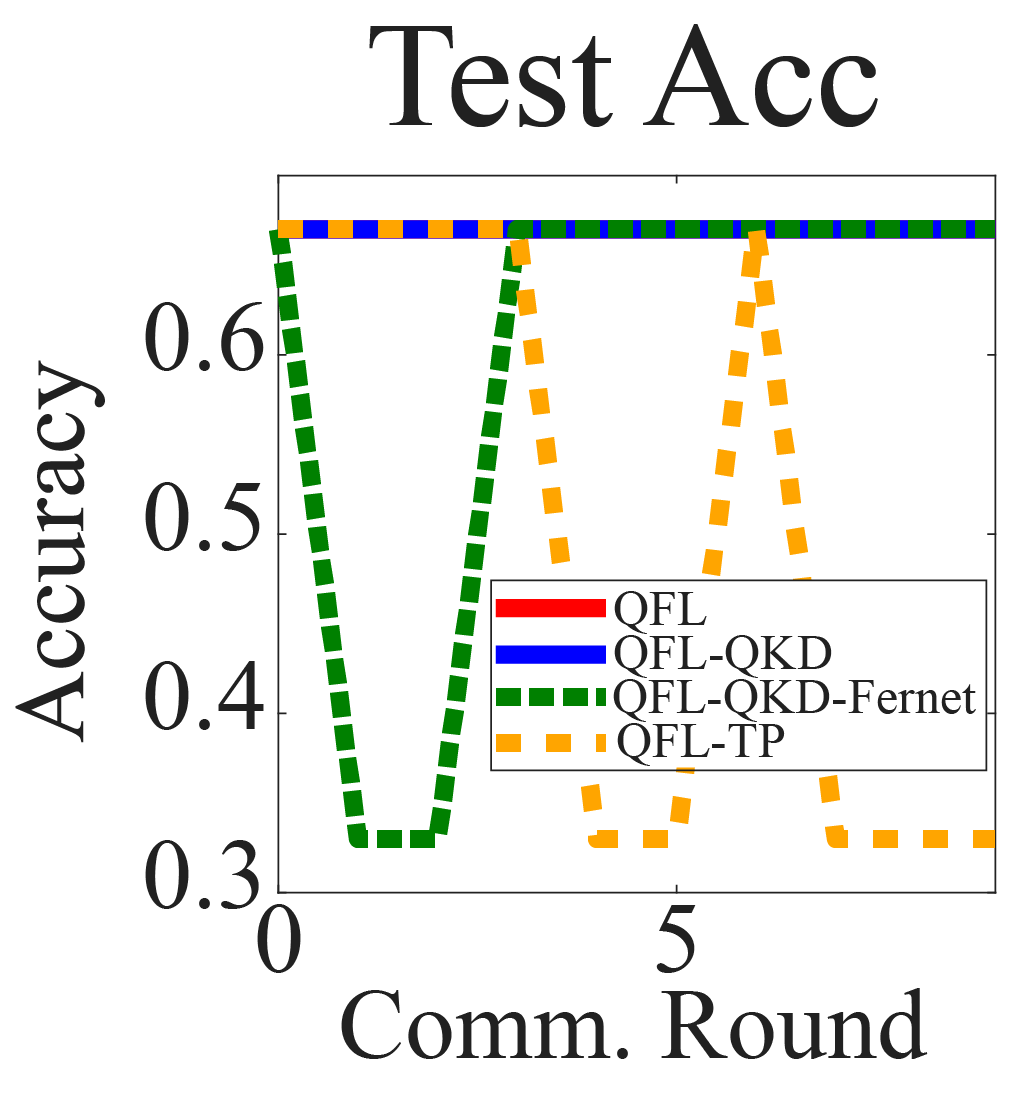}
    \caption{Server}
    \label{fig:server_test_acc_qkd_fernet_tp}
    \end{subfigure}
      \begin{subfigure}[b]{0.32\columnwidth}
        \centering
       \includegraphics[width=\columnwidth]{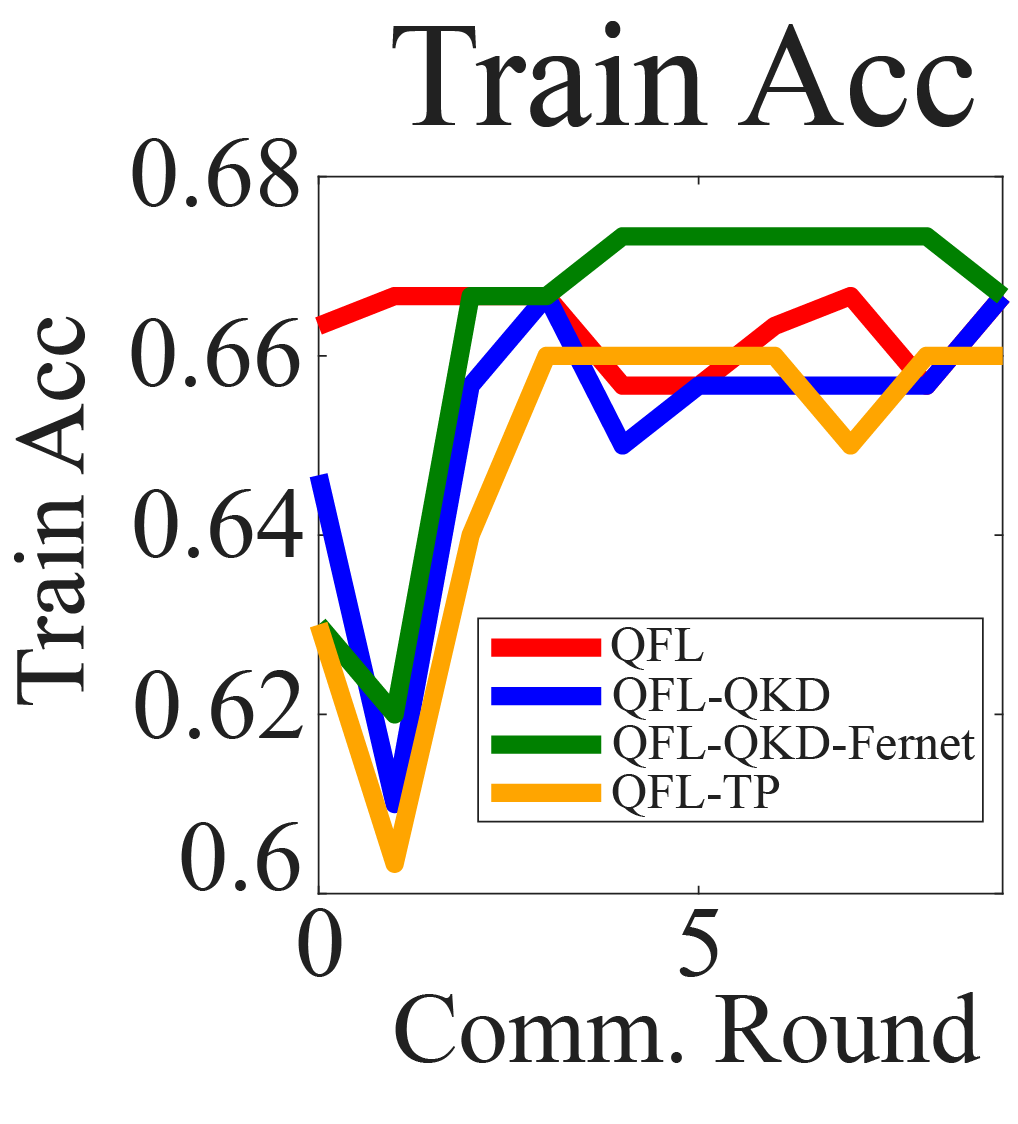}
    \caption{Devices}
    \label{fig:avg_devices_val_acc_qkd_fernet_tp}
    \end{subfigure}
      \begin{subfigure}[b]{0.32\columnwidth}
        \centering
       \includegraphics[width=\linewidth]{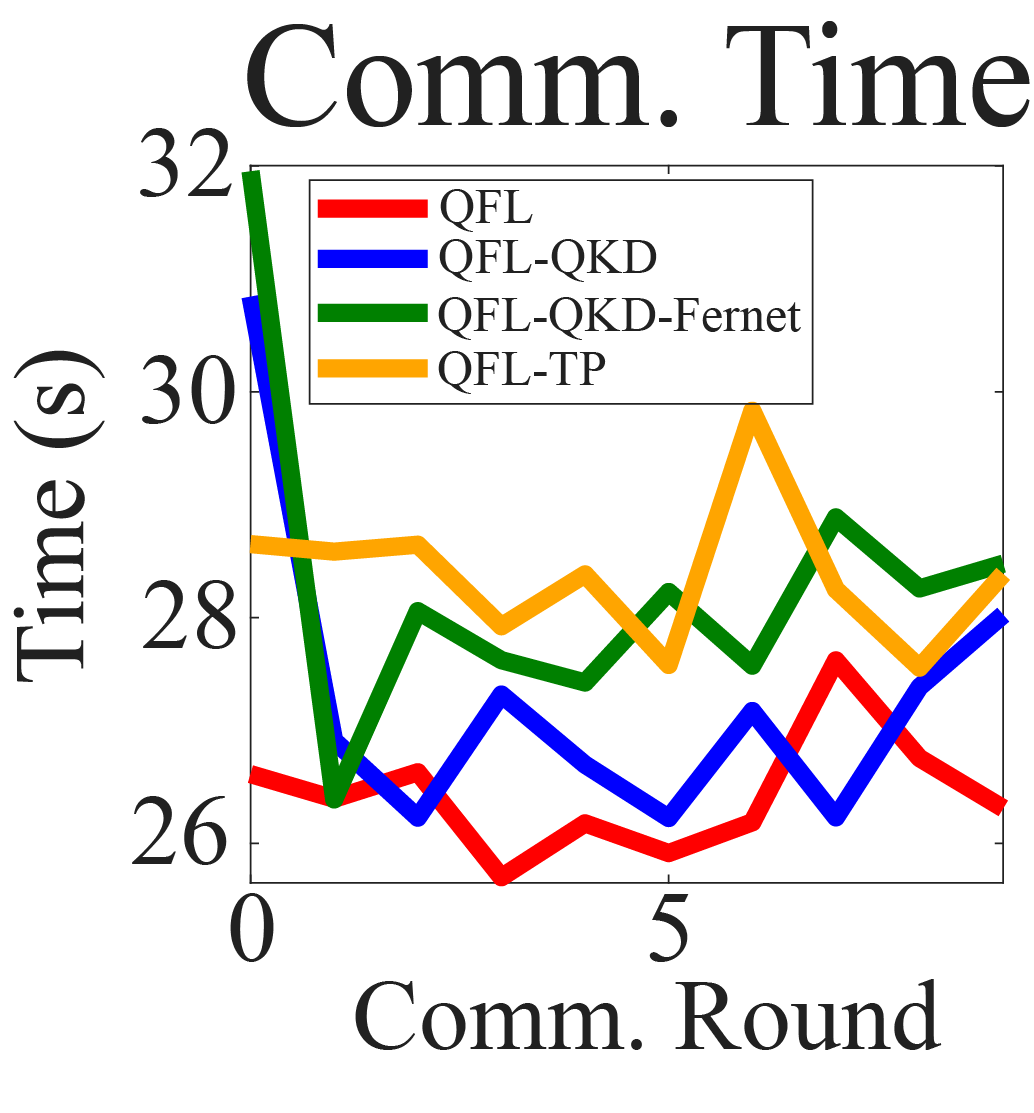}
    \caption{Comm Time}
    \label{fig:comm_time_qkd_fernet_tp}
    \end{subfigure}   
    \caption{Overall Performance: QFL, QFL-QKD, QFL-QKD-Fernet, and QFL-TP}
    \label{fig:server_performance_qkd_fernet_tp}
\end{figure}

Figure \ref{fig:server_performance_qkd_fernet_tp} shows the comparative results of QFL frameworks with various implementations of protocols: standard QFL, QFL-QKD, QFL-QKD-Fernet, and QFL-TP.
As mentioned above, these are only observation results showing slight variations on the performance of those frameworks when QKD, QKD-Fernet or QFL-TP is integrated.
However, noticeable differences can be observed in terms of communication time, which shows that QFL is the fastest as expected with TL integration slightly slowing down the process. 
From this we can conclude that the actual real-world implementations of these protocols will have huge impact on overall computational overhead.

\subsubsection{KEM}
\begin{figure}[!htbp]
    \centering
    \begin{subfigure}[b]{0.3\columnwidth}
        \centering
       \includegraphics[width=\columnwidth]{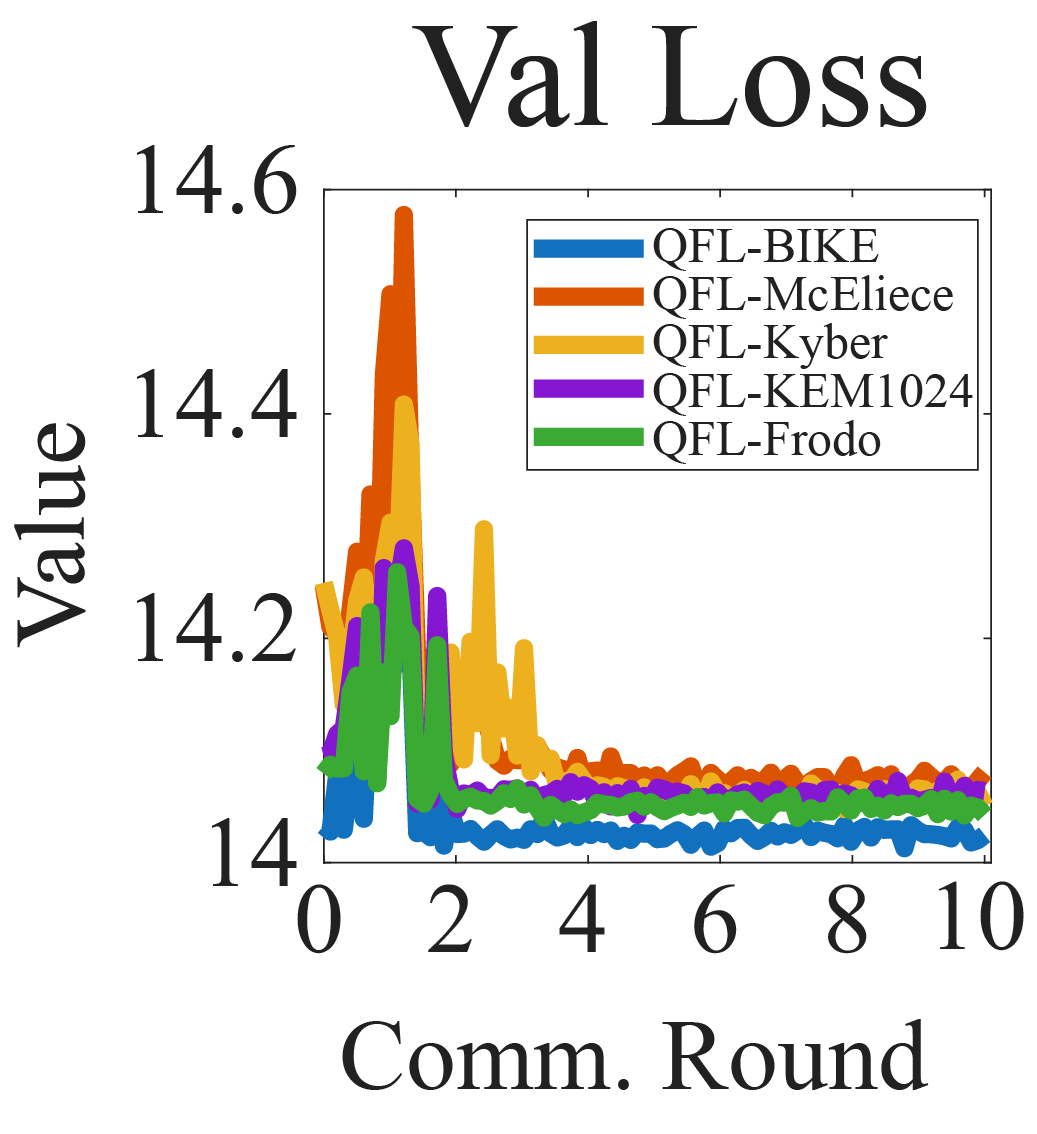}
    \caption{Server Val Loss}
    \label{fig:server_val_loss_kem}
    \end{subfigure}
    \begin{subfigure}[b]{0.29\columnwidth}
        \centering
       \includegraphics[width=\columnwidth]{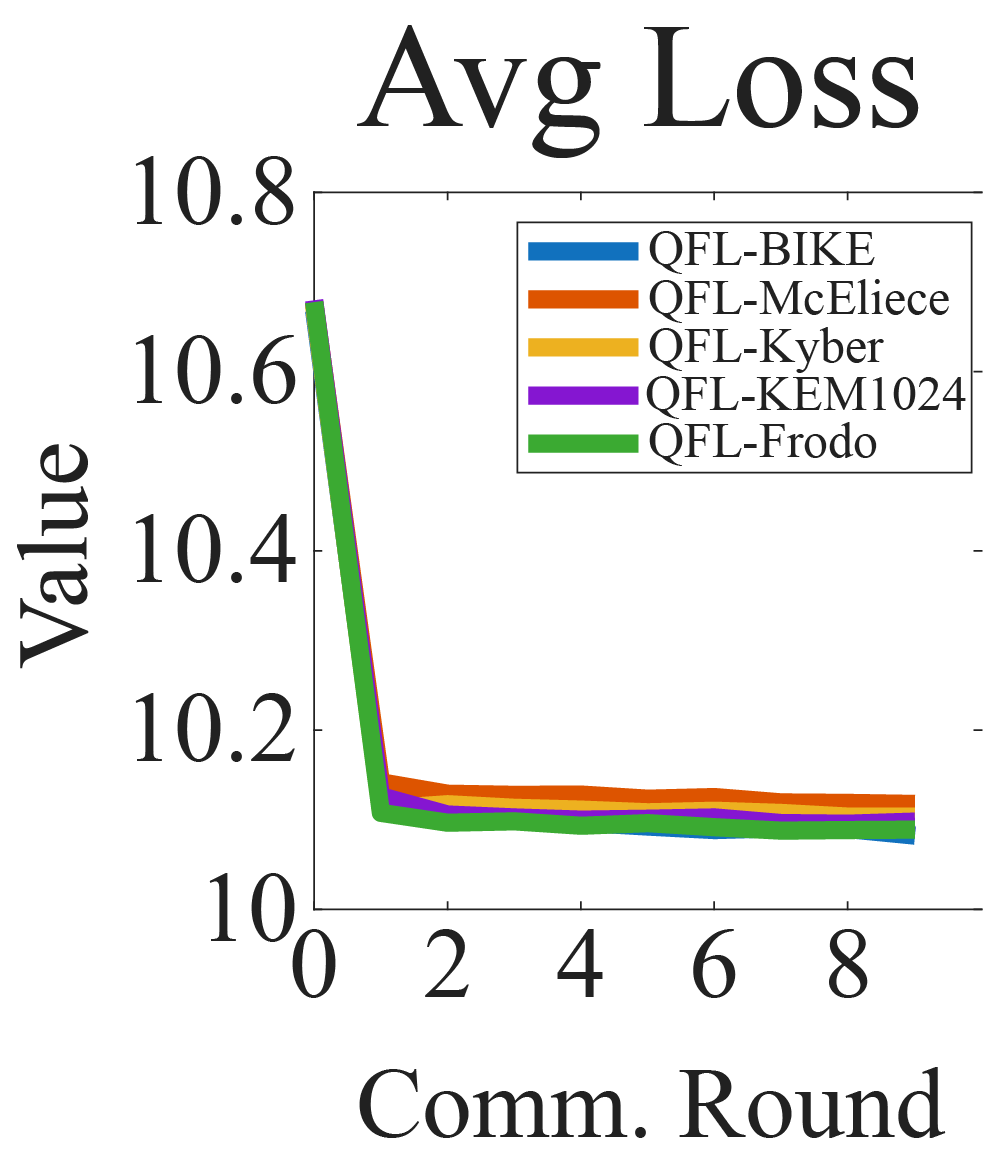}
    \caption{Train Loss}
    \label{fig:avg_devices_obj_values_kem}
    \end{subfigure}
  \begin{subfigure}[b]{0.3\columnwidth}
        \centering
      \includegraphics[width=\columnwidth]{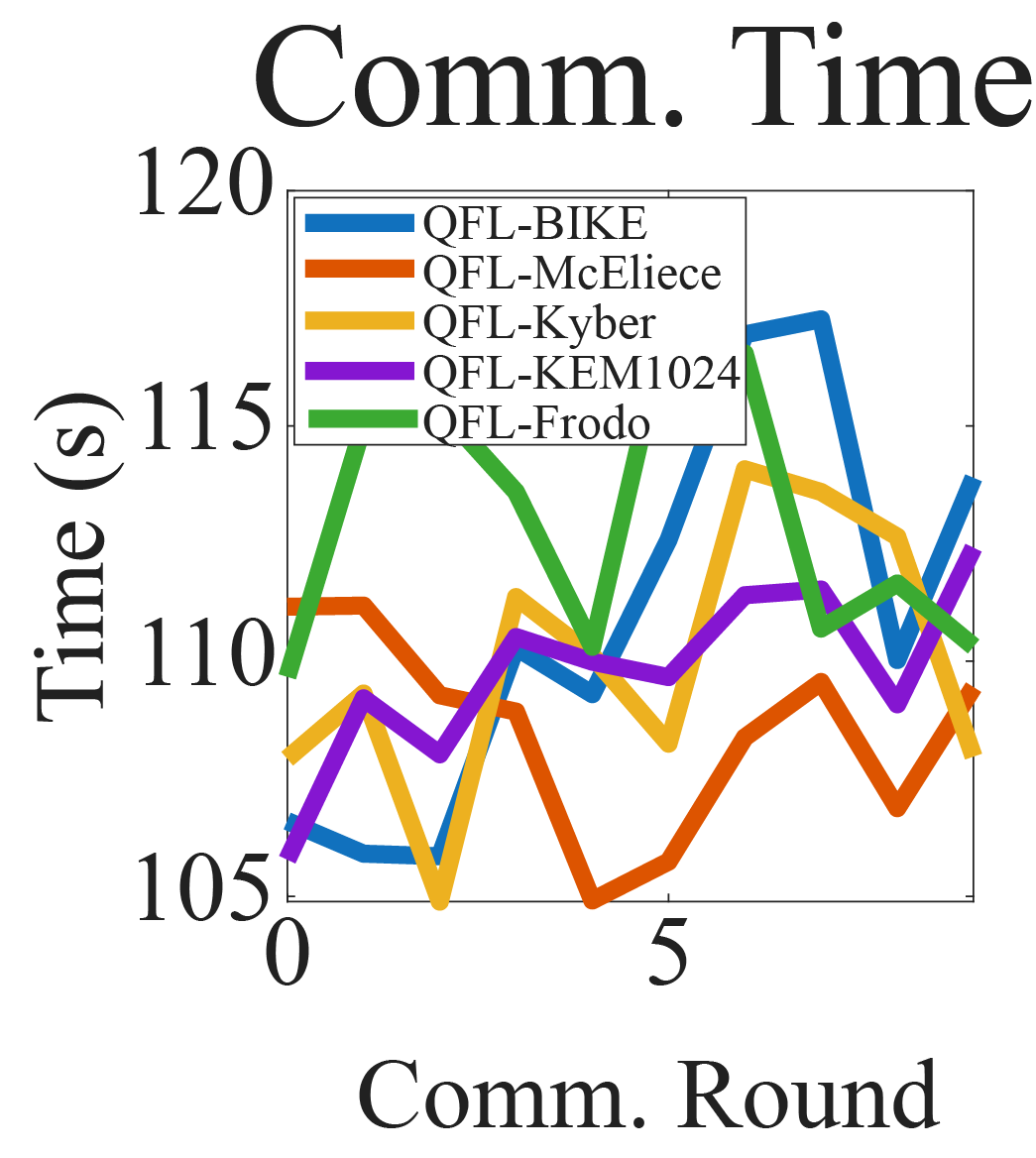}
    \caption{Comm Time}
    \label{fig:comm_time_kem}
    \end{subfigure}
    \caption{Overall Performance: KEM Results}
    \label{fig:overall_performance_kem}
\end{figure}

Key Encapsulation mechanism is primarily used to encapsulate the private key to share between two parties.
In this study, we introduced the concept of using KEM keys to secure model parameters using hash values. 
This represents an initial exploration, necessitating more comprehensive investigations to fully understand the practical implementation requirements and considerations.
The KEM mechanisms compared are BIKE, McEliece, Kyber, KEM1024, FrodoKEM etc.
The use of different mechanisms has resulted in varying results, as seen in Figure \ref{fig:overall_performance_kem}, with QFL-Bike being the slowest in terms of communication time. 
It shows that each scheme is different, and thus various considerations are required to be made and understood for their implementations.

\subsubsection{PQC}
In terms of implementation with PQC schemes including Dilithium, Falcon1024, ML-DSA024, SPHINCS, Mayo etc., the results are varying with SPHINCS being the slowest of all with Falcon1024 the fastest.
Since each PQC schemes are made differently, they all provide different unique advantages with their own computational limitations such as key generation, signing etc. 
The detailed comparative results among various PQC schemes and KEMs are presented in the appendix.

\begin{figure}[!htbp]
    \centering
    \begin{subfigure}[b]{0.32\columnwidth}
        \centering
       \includegraphics[width=\columnwidth]{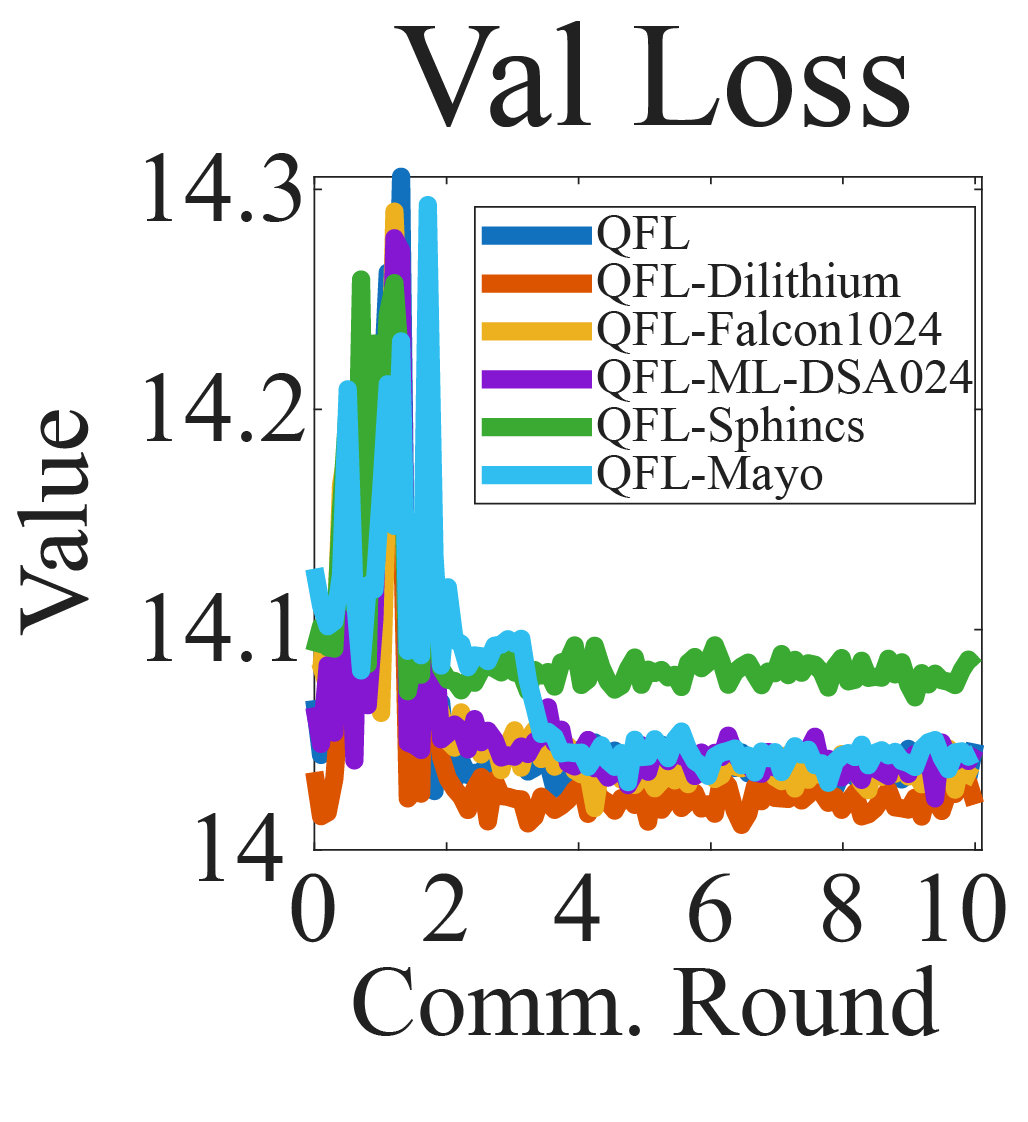}
    \caption{Server Val Loss}
    \label{fig:server_obj_values_pqc}
    \end{subfigure}
    \begin{subfigure}[b]{0.32\columnwidth}
        \centering
       \includegraphics[width=\columnwidth]{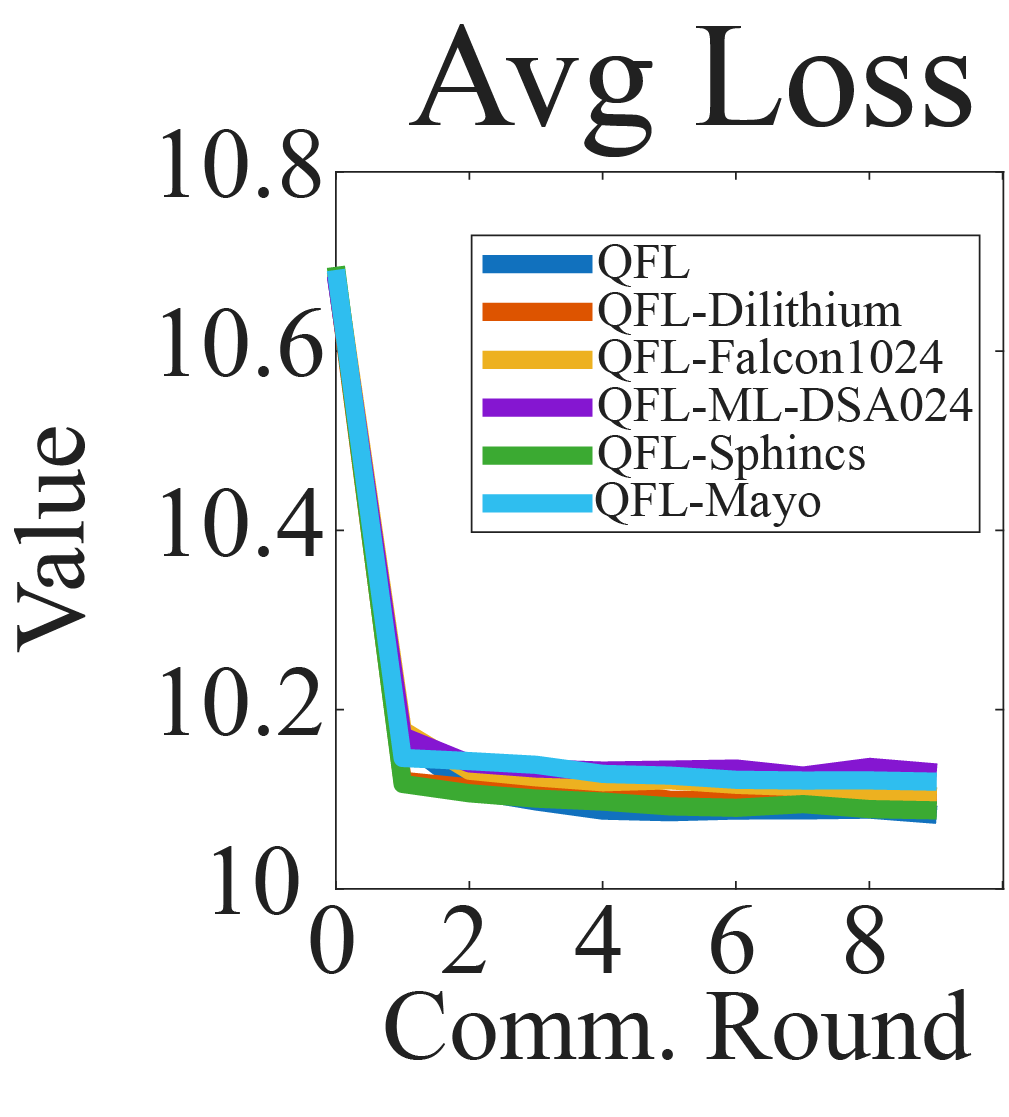}
    \caption{Avg Devices Loss}
    \label{fig:avg_devices_obj_values_pqc}
    \end{subfigure}
  \begin{subfigure}[b]{0.32\columnwidth}
        \centering
      \includegraphics[width=\columnwidth]{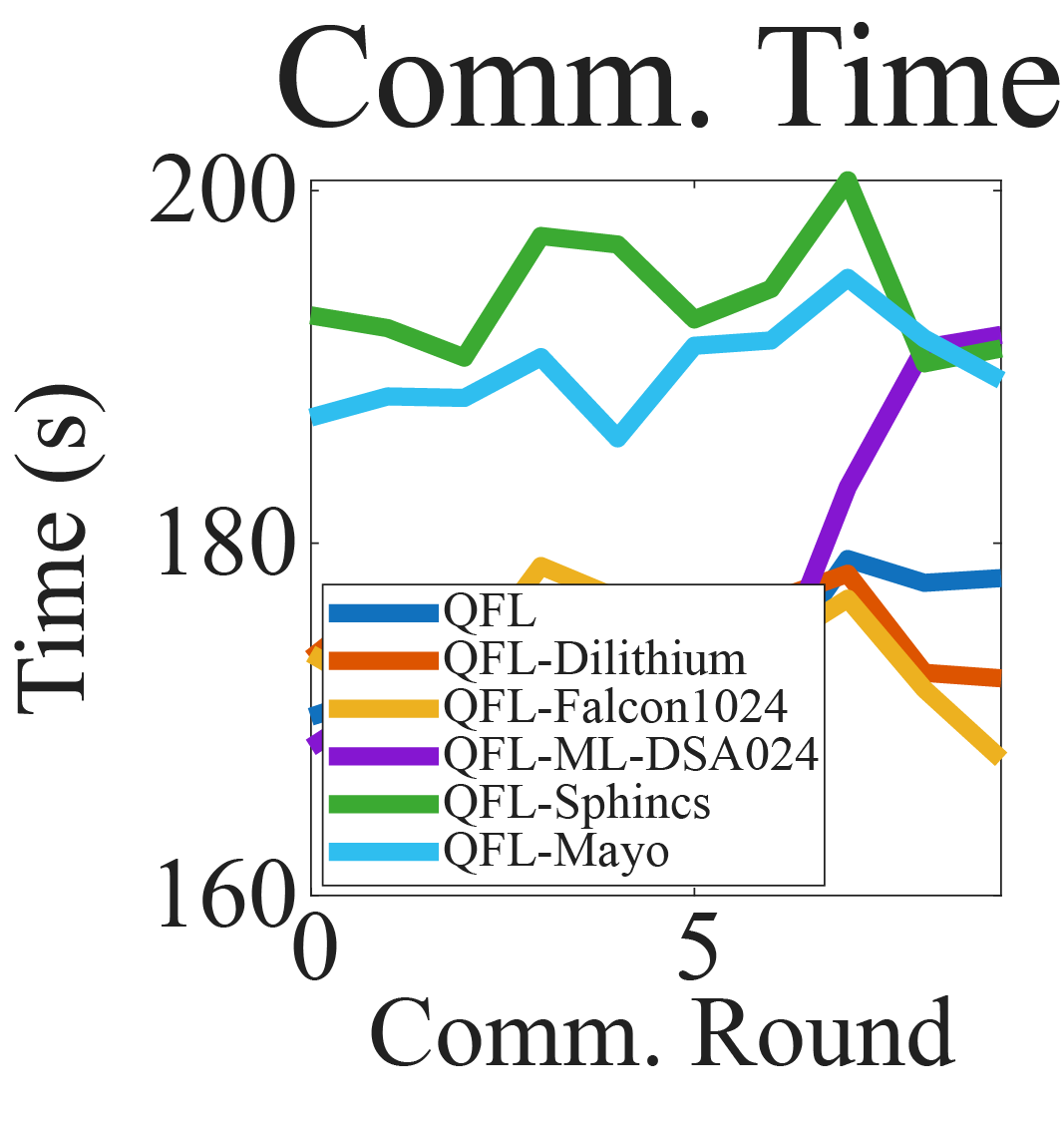}
    \caption{Comm Time}
    \label{fig:comm_time_pqc}
    \end{subfigure}
    \caption{PQC Schemes Implementation}
    \label{fig:server_performance}
\end{figure}

\subsubsection{Dataset}
Table \ref{tab:genomic_vs_iris_dataset_results} shows varying results depending on the type of dataset used.
Genomic is a two class label dataset whereas IRIS is a three class label dataset.
We can see in terms of accuracy and other factors that performance is better on Genomic Dataset.
This highlights the capability and limitations of quantum classifiers like VQC.

\begin{table}[h]
\centering
\caption{IRIS and Genomic Dataset}
\label{tab:genomic_vs_iris_dataset_results}
\resizebox{\columnwidth}{!}{
\begin{tabular}{l l p{0.7cm} p{0.7cm} p{0.7cm} p{0.7cm} p{0.8cm} p{0.8cm} p{1cm}}
\toprule
Dataset & Model & Avg Val Acc & Final Val Acc & Avg Test Acc & Final Test Acc & Avg Val Loss & Final Val Loss & Avg Comm Time (s) \\
\midrule
\multirow{4}{*}{Iris} 
    & QFL & 0.572 & 0.5 & 0.67 & 0.67 & 14.14 & 14.21 & 26.43 \\
    & QFL-QKD & 0.58 & 0.58 & 0.67 & 0.67 & 14.14 & 14.15 & 27.3 \\
    & QFL-QKD\_Fernet & 0.556 & 0.58 & 0.602 & 0.67 & 14.13 & 14.12 & 28.29 \\
    & QFL-TP & 0.532 & 0.5 & 0.5 & 0.33 & 14.13 & 14.17 & 28.38 \\
\midrule
\multirow{4}{*}{Genomic} 
    & QFL & 0.5915 & 0.57 & 0.739 & 0.73 & 0.95 & 0.99 & 211.56 \\
    & QFL-QKD & \textbf{0.6335} & 0.65 & 0.5715 & 0.57 & \textbf{0.93} & \textbf{0.96} & 213.01 \\
    & QFL-QKD\_Fernet & 0.6045 & \textbf{0.61} & 0.585 & 0.5 & 0.94 & 0.98 & 211.64 \\
    & QFL-TP & 0.609 & 0.57 & 0.643 & 0.63 & 0.94 & 0.97 & 212.8 \\
\bottomrule
\end{tabular}
}
\end{table}



\section{Conclusion and Future Work}
In this study, we introduced a robust quantum federated learning (QFL) framework, utilizing quantum key distribution, teleportation, 
key encapsulation mechanisms (KEM), and post-quantum cryptography (PQC) methods. 
We outlined a range of algorithms and protocols to 
facilitate the practical application of these technologies within 
the domain of QFL, with a particular focus on secure quantum communication. 
Our experimental findings, along with theoretical analyses and observations, 
are poised to serve as a foundational basis for 
future research in the domain of secure QFL frameworks.

There are various fields in which this study can be extended further.
First, the study of photon attacks is promising, such as the study of BB84 decoy state for better security \cite{tupkaryQKDSecurityProofs2025}. 
Similarly, study of quantum networks especially entangled distribution to achieve quantum internet for multi-party QKD can be crucial \cite{cacciapuotiEntanglementDistributionQuantum2024}.
Another area of study is the need for homomorphic encryption for aggregated computation. 
But the challenge is still there, since most of these studies highly depend on progress in the field of quantum computing regarding the actual design of a fault-tolerant quantum computer.




\clearpage
\section{Appendix}
\subsection{Classical Vs PQC Schemes}
\begin{figure*}[!htbp]
    \centering
    \includegraphics[width=\linewidth]{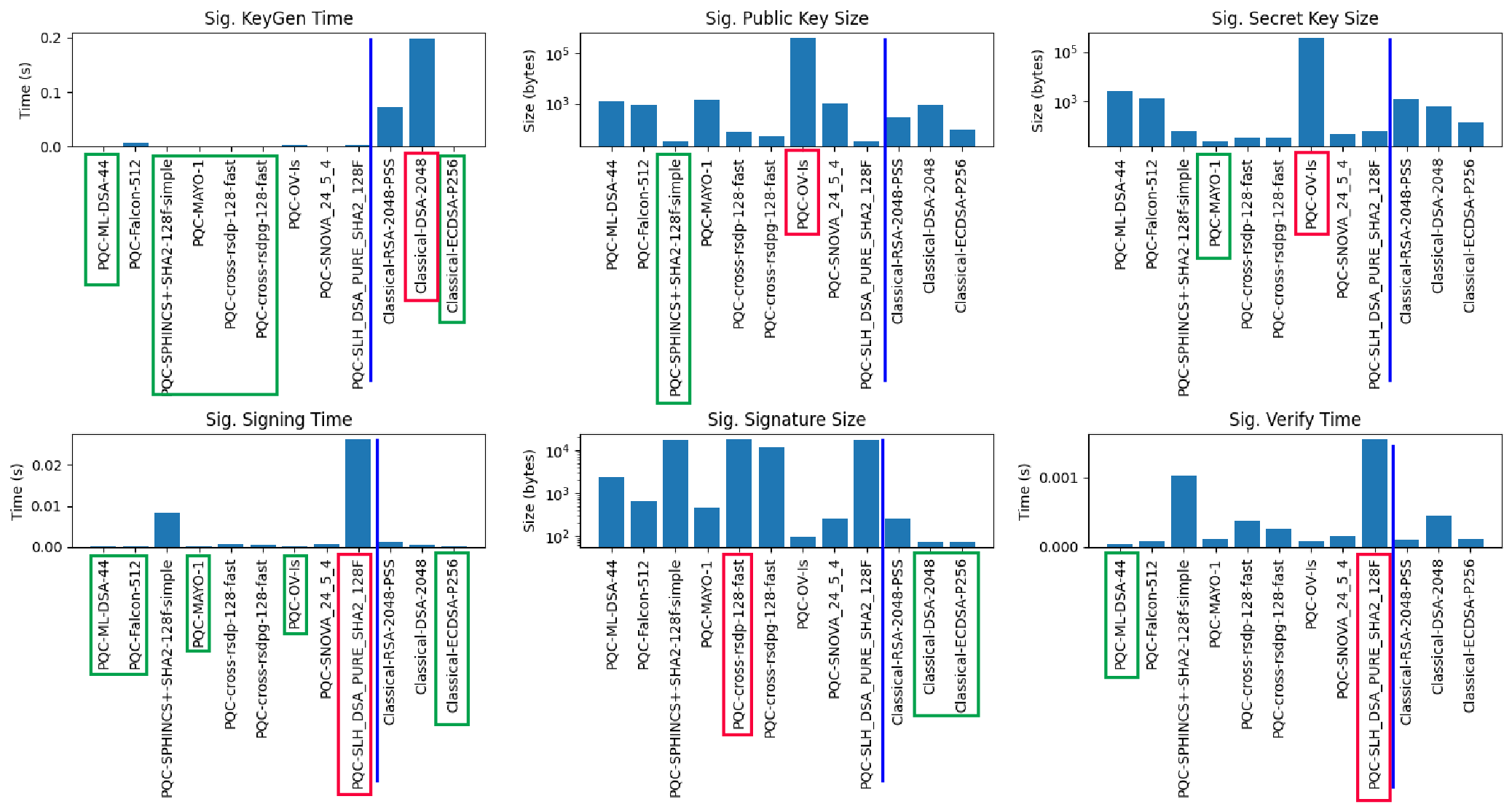}
    \caption{PQC Quantum vs Classical}
    \label{fig:public_key_quantum_vs_classica}
\end{figure*}

\begin{figure*}[!htbp]
    \centering
    \includegraphics[width=\linewidth]{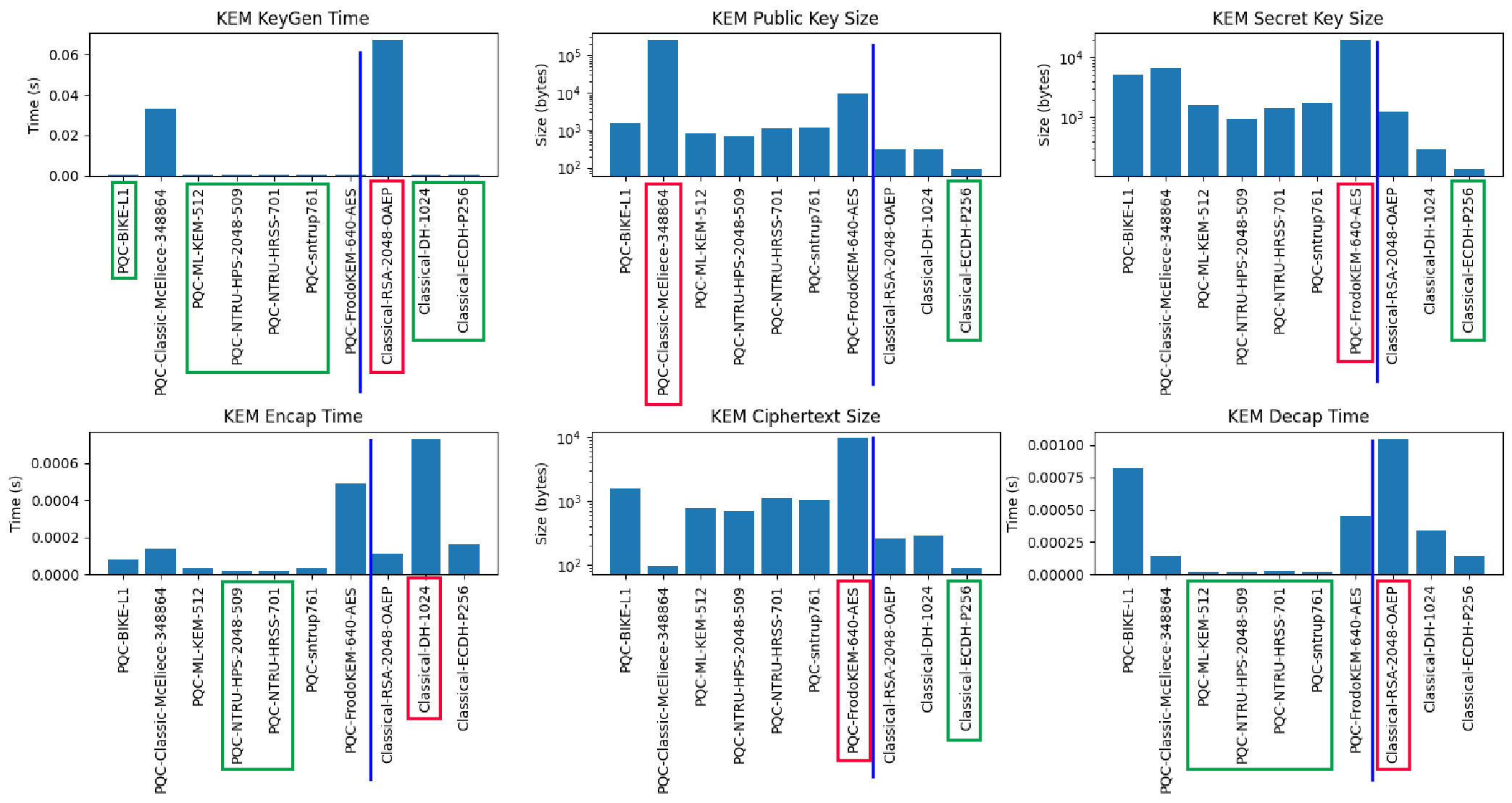}
    \caption{KEM Quantum vs Classical}
    \label{fig:kem_quantum_vs_classical}
\end{figure*}

In Figures \ref{fig:public_key_quantum_vs_classica} and \ref{fig:kem_quantum_vs_classical}, we can observe differences in the generation of signature keys, the signing and verification time, the size of public, private, and signature keys.
In terms of public key types, 
classical cryptography like RSA, DSA and ECDSA is better with signing time similar to with PQC schemes such as DSA, Falcon etc. 
It should be noted that the time required for key generation in traditional RSA is slower compared to post-quantum cryptography (PQC) schemes such as DSA and Falcon, among others.
However, in terms of KEM schemes as in Figure \ref{fig:kem_quantum_vs_classical}, the
classical DH and ECDH algorithms are comparable to BIKE, ML-KEM, etc.  in terms of key generation time.
The good advantage with classical cryptography is seen in terms of public-key sizes and secret-key sizes.
However, in terms of encapsulation time, PQC schemes such as ML-KEM and HPS are better.
This shows that each cryptography scheme is good and bad in their own ways.
However, considering quantum security, the only option is PQC schemes.

\subsection{Signature Schemes}
\begin{figure}[!htbp]
     \centering
     \begin{subfigure}[b]{\columnwidth}
         \centering
         \includegraphics[width=\columnwidth]{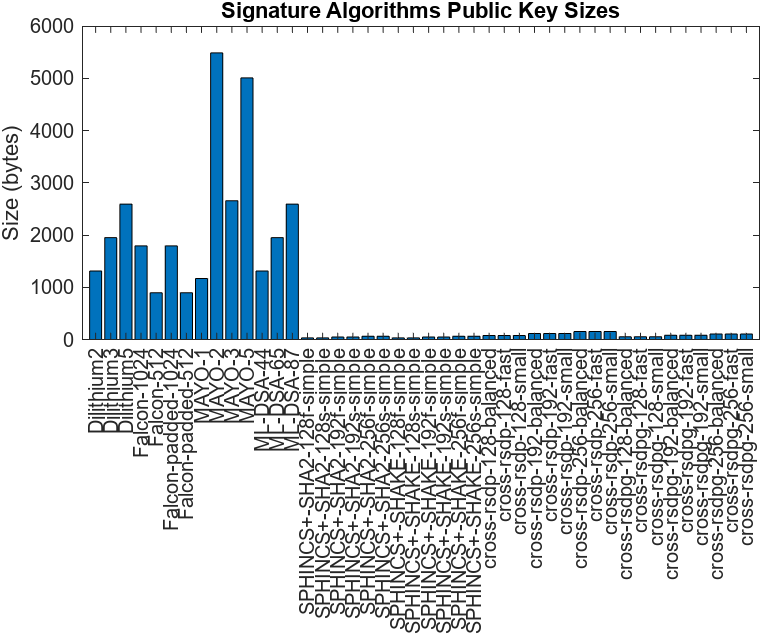}
         \caption{Public Key Size}
         \label{fig:pub_keys_sig_schemes}
     \end{subfigure}
     \hfill
     \begin{subfigure}[b]{\columnwidth}
         \centering
         \includegraphics[width=\columnwidth]{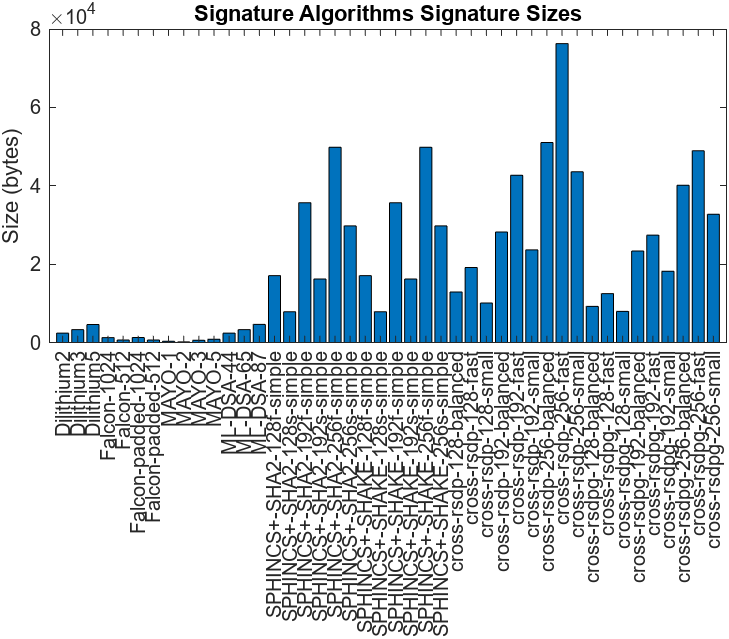}
         \caption{Signature Size}
         \label{fig:signature_sizes}
     \end{subfigure}
     \hfill
     \begin{subfigure}[b]{\columnwidth}
         \centering
         \includegraphics[width=\columnwidth]{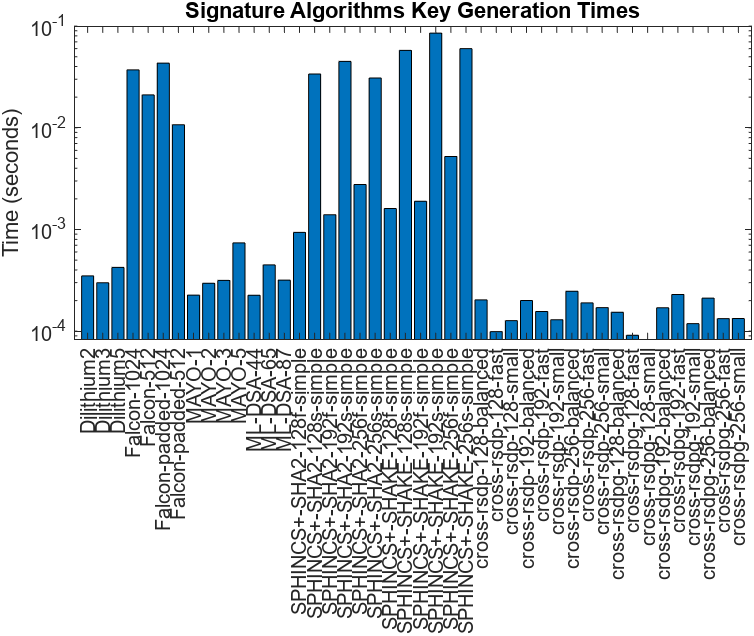}
         \caption{Key Generation Time}
         \label{fig:key_gen_time_sig}
     \end{subfigure}
        \caption{Signature Schemes}
        \label{fig:signature_schemes}
\end{figure}

The bar graph in Figure \ref{fig:pub_keys_sig_schemes} illustrates the public key sizes (in bytes) for different signature algorithms.
Most schemes have key sizes ranging from 1,000 to 5,000 bytes.
Public key size is a crucial aspect in the environment where multiple key generation and signing and transmitting is required.
Thus, schemes like SPHINCS+ having small public key sizes is very beneficial.
However, in terms of signature size, as shown in Figure \ref{fig:signature_sizes}, schemes having larger public sizes seem to have smaller signature sizes. 
Thus, there is a clear trade-off. 
These schemes all provide quantum security and thus can be chosen based on the settings where either signature size is important or public key sizes.
Similarly, another crucial component is the key generation time, which is shown in Figure \ref{fig:key_gen_time_sig}.
In this category, CROSS schemes seem to be the fastest signature schemes.

\subsection{KEM Schemes}
\begin{figure}[!htbp]
     \centering
     \begin{subfigure}[b]{\columnwidth}
         \centering
         \includegraphics[width=\columnwidth]{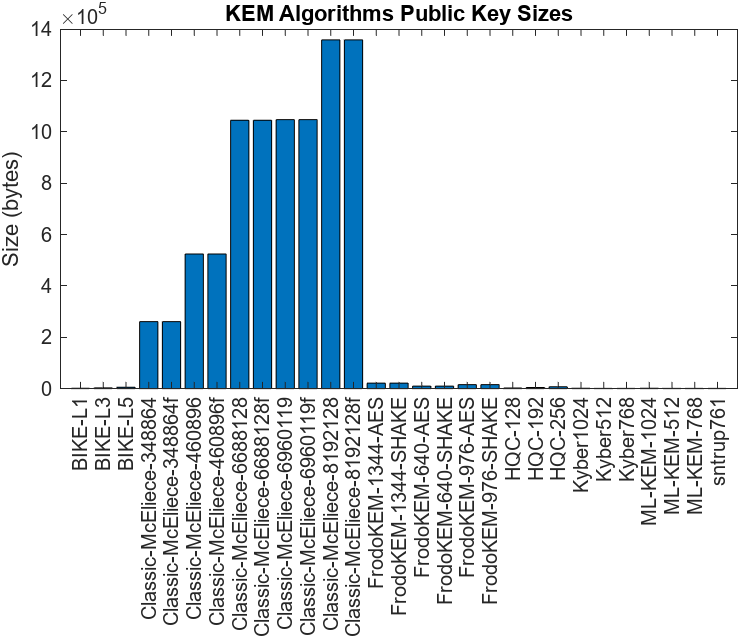}
         \caption{Public Key Size}
         \label{fig:pub_keys_size_kem}
     \end{subfigure}
     \hfill
     \begin{subfigure}[b]{\columnwidth}
         \centering
         \includegraphics[width=\columnwidth]{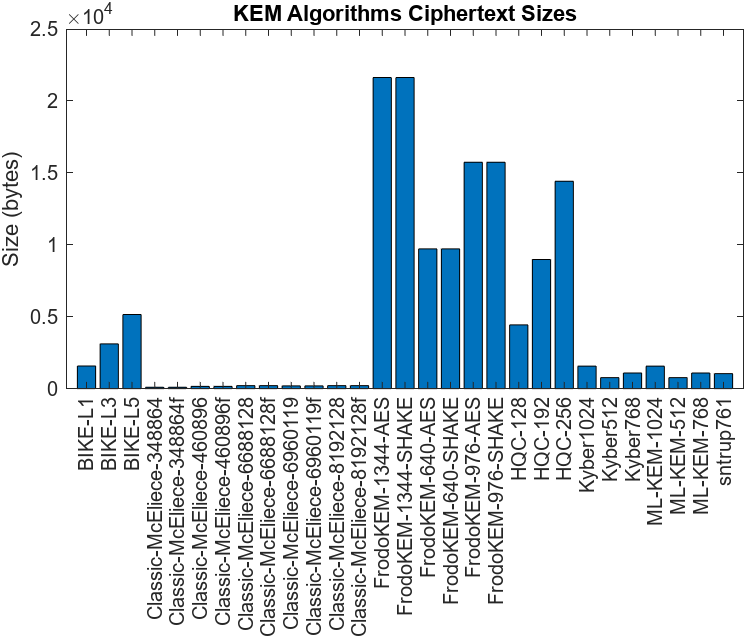}
         \caption{Cipher Text Size}
         \label{fig:ciphertext_size_kem}
     \end{subfigure}
     \hfill
     \begin{subfigure}[b]{\columnwidth}
         \centering
         \includegraphics[width=\columnwidth]{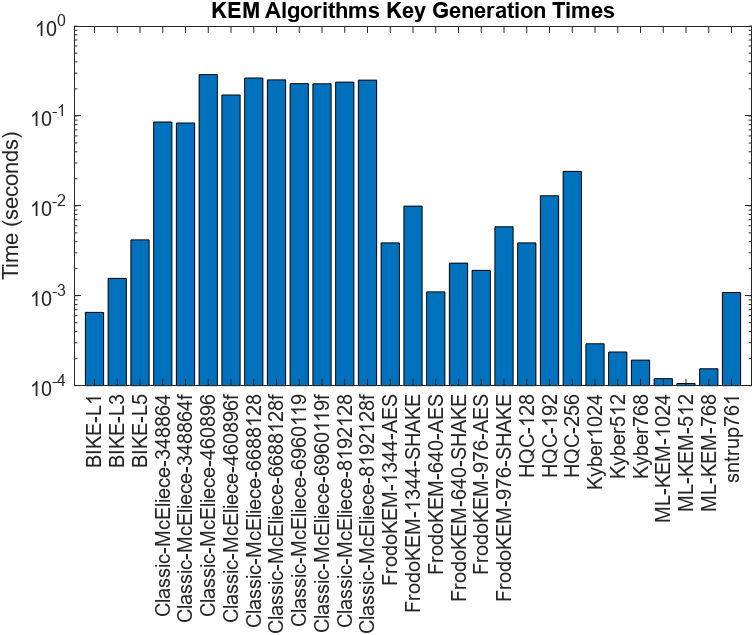}
         \caption{Key Generation Time}
         \label{fig:key_generation_time_kem}
     \end{subfigure}
        \caption{KEM Schemes}
        \label{fig:kem_schemes}
\end{figure}

In Figure \ref{fig:kem_schemes}, we compare the public key size, the ciphertext size, and the key generation time of various KEM schemes such as BIKE, McEliece, FrodoKEM, HQC, Kyber, ML-KEM, etc.
In Figure \ref{fig:pub_keys_size_kem}, we can see that public key sizes for Kyber, BIKE, ML-KEM etc. are way smaller than that of schemes like McElience etc. 
This shows how different each KEM schemes are and thus are suited for different situations.
This also concludes that one scheme that is best in every aspect is hard to find.
Because, now in terms of CipherText sizes, we can see that Classic-McEliece is much smaller than others.
However, in terms of key generation time, we see that ML-KEM is the fastest KEM scheme.

\subsection{Table Summary}
In Table \ref{tab:pqc_signature_schemes}, we present the simulation results in the Google colab in terms of Keygen times, PK sizes, SK sizes, sign times, sig sizes, and verification times.
We can see that each signature scheme has both good and bad factors.
In terms of KeyGen times, we can see that MAYO and CROSS are better.
However, in terms of PK sizes, MAYO is the worst.
In terms of signing time, Dilithium and ML-DSA are good.
For signatures sizes, Falcon and MAYO are better with verification time with Dilithium, Falcon, SPHINCS+ etc.
Thus, we can see that a single signature scheme is not perfect in all aspects.
Thus, the selection of signature schemes needs to be done on the basis of the need and scenario.

\begin{table}[]
\centering
\resizebox{\columnwidth}{!}{
\begin{tabular}{lcccccc}
\toprule
Schemes & Keygen Times & PK sizes & SK sizes & Sign Times & Sig Sizes & Verify Times \\
\midrule
Dilithium2 & \cellcolor{yellow!50}{0.000350} & \cellcolor{green!50}{1312} & \cellcolor{yellow!50}{2528} & \cellcolor{green!50}{0.000280} & \cellcolor{yellow!50}{2420} & \cellcolor{green!50}{0.000087} \\
Dilithium3 & \cellcolor{green!50}{0.000299} & \cellcolor{yellow!50}{1952} & \cellcolor{orange!50}{4000} & \cellcolor{green!50}{0.000272} & \cellcolor{yellow!50}{3293} & \cellcolor{yellow!50}{0.000126} \\
Dilithium5 & \cellcolor{yellow!50}{0.000423} & \cellcolor{orange!50}{2592} & \cellcolor{red!50}{4864} & \cellcolor{green!50}{0.000323} & \cellcolor{orange!50}{4595} & \cellcolor{yellow!50}{0.000177} \\
ML-DSA-44 & \cellcolor{green!50}{0.000225} & \cellcolor{green!50}{1312} & \cellcolor{yellow!50}{2560} & \cellcolor{green!50}{0.000169} & \cellcolor{yellow!50}{2420} & \cellcolor{yellow!50}{0.000136} \\
ML-DSA-65 & \cellcolor{yellow!50}{0.000448} & \cellcolor{yellow!50}{1952} & \cellcolor{orange!50}{4032} & \cellcolor{green!50}{0.000333} & \cellcolor{yellow!50}{3309} & \cellcolor{green!50}{0.000122} \\
ML-DSA-87 & \cellcolor{green!50}{0.000318} & \cellcolor{orange!50}{2592} & \cellcolor{red!50}{4896} & \cellcolor{yellow!50}{0.000872} & \cellcolor{orange!50}{4627} & \cellcolor{yellow!50}{0.000191} \\
Falcon-512 & \cellcolor{red!50}{0.021049} & \cellcolor{green!50}{897} & \cellcolor{green!50}{1281} & \cellcolor{orange!50}{0.001708} & \cellcolor{green!50}{656} & \cellcolor{yellow!50}{0.000182} \\
Falcon-1024 & \cellcolor{red!50}{0.037076} & \cellcolor{yellow!50}{1793} & \cellcolor{yellow!50}{2305} & \cellcolor{orange!50}{0.001764} & \cellcolor{green!50}{1270} & \cellcolor{orange!50}{0.000322} \\
Falcon-padded-512 & \cellcolor{orange!50}{0.010720} & \cellcolor{green!50}{897} & \cellcolor{green!50}{1281} & \cellcolor{yellow!50}{0.000567} & \cellcolor{green!50}{666} & \cellcolor{green!50}{0.000087} \\
Falcon-padded-1024 & \cellcolor{red!50}{0.043175} & \cellcolor{yellow!50}{1793} & \cellcolor{yellow!50}{2305} & \cellcolor{yellow!50}{0.000945} & \cellcolor{green!50}{1280} & \cellcolor{green!50}{0.000159} \\
SPHINCS+-SHA2-128f-simple & \cellcolor{yellow!50}{0.000936} & \cellcolor{green!50}{32} & \cellcolor{green!50}{64} & \cellcolor{red!50}{0.013423} & \cellcolor{red!50}{17088} & \cellcolor{orange!50}{0.001133} \\
SPHINCS+-SHA2-128s-simple & \cellcolor{red!50}{0.033746} & \cellcolor{green!50}{32} & \cellcolor{green!50}{64} & \cellcolor{red!50}{0.219949} & \cellcolor{orange!50}{7856} & \cellcolor{green!50}{0.000391} \\
SPHINCS+-SHA2-192f-simple & \cellcolor{yellow!50}{0.001396} & \cellcolor{green!50}{48} & \cellcolor{green!50}{96} & \cellcolor{red!50}{0.022754} & \cellcolor{red!50}{35664} & \cellcolor{orange!50}{0.001352} \\
SPHINCS+-SHA2-192s-simple & \cellcolor{red!50}{0.044950} & \cellcolor{green!50}{48} & \cellcolor{green!50}{96} & \cellcolor{red!50}{0.423860} & \cellcolor{red!50}{16224} & \cellcolor{yellow!50}{0.000899} \\
SPHINCS+-SHA2-256f-simple & \cellcolor{orange!50}{0.002767} & \cellcolor{green!50}{64} & \cellcolor{green!50}{128} & \cellcolor{red!50}{0.039144} & \cellcolor{red!50}{49856} & \cellcolor{orange!50}{0.001346} \\
SPHINCS+-SHA2-256s-simple & \cellcolor{red!50}{0.030869} & \cellcolor{green!50}{64} & \cellcolor{green!50}{128} & \cellcolor{red!50}{0.374365} & \cellcolor{red!50}{29792} & \cellcolor{green!50}{0.000788} \\
SPHINCS+-SHAKE-128f-simple & \cellcolor{yellow!50}{0.001608} & \cellcolor{green!50}{32} & \cellcolor{green!50}{64} & \cellcolor{red!50}{0.025012} & \cellcolor{red!50}{17088} & \cellcolor{orange!50}{0.001478} \\
SPHINCS+-SHAKE-128s-simple & \cellcolor{red!50}{0.057624} & \cellcolor{green!50}{32} & \cellcolor{green!50}{64} & \cellcolor{red!50}{0.443206} & \cellcolor{orange!50}{7856} & \cellcolor{green!50}{0.000589} \\
SPHINCS+-SHAKE-192f-simple & \cellcolor{yellow!50}{0.001894} & \cellcolor{green!50}{48} & \cellcolor{green!50}{96} & \cellcolor{red!50}{0.034911} & \cellcolor{red!50}{35664} & \cellcolor{red!50}{0.002084} \\
SPHINCS+-SHAKE-192s-simple & \cellcolor{red!50}{0.085185} & \cellcolor{green!50}{48} & \cellcolor{green!50}{96} & \cellcolor{red!50}{0.738338} & \cellcolor{red!50}{16224} & \cellcolor{green!50}{0.000749} \\
SPHINCS+-SHAKE-256f-simple & \cellcolor{orange!50}{0.005212} & \cellcolor{green!50}{64} & \cellcolor{green!50}{128} & \cellcolor{red!50}{0.088743} & \cellcolor{red!50}{49856} & \cellcolor{red!50}{0.002458} \\
SPHINCS+-SHAKE-256s-simple & \cellcolor{red!50}{0.060087} & \cellcolor{green!50}{64} & \cellcolor{green!50}{128} & \cellcolor{red!50}{0.649662} & \cellcolor{red!50}{29792} & \cellcolor{orange!50}{0.001077} \\
MAYO-1 & \cellcolor{green!50}{0.000226} & \cellcolor{green!50}{1168} & \cellcolor{green!50}{24} & \cellcolor{green!50}{0.000254} & \cellcolor{green!50}{321} & \cellcolor{green!50}{0.000071} \\
MAYO-2 & \cellcolor{green!50}{0.000296} & \cellcolor{red!50}{5488} & \cellcolor{green!50}{24} & \cellcolor{green!50}{0.000323} & \cellcolor{green!50}{180} & \cellcolor{green!50}{0.000086} \\
MAYO-3 & \cellcolor{green!50}{0.000316} & \cellcolor{orange!50}{2656} & \cellcolor{green!50}{32} & \cellcolor{yellow!50}{0.000719} & \cellcolor{green!50}{577} & \cellcolor{yellow!50}{0.000203} \\
MAYO-5 & \cellcolor{yellow!50}{0.000739} & \cellcolor{red!50}{5008} & \cellcolor{green!50}{40} & \cellcolor{orange!50}{0.002314} & \cellcolor{green!50}{838} & \cellcolor{orange!50}{0.000362} \\
cross-rsdp-128-balanced & \cellcolor{green!50}{0.000203} & \cellcolor{green!50}{77} & \cellcolor{green!50}{32} & \cellcolor{orange!50}{0.001732} & \cellcolor{orange!50}{12912} & \cellcolor{orange!50}{0.001286} \\
cross-rsdp-128-fast & \cellcolor{green!50}{0.000099} & \cellcolor{green!50}{77} & \cellcolor{green!50}{32} & \cellcolor{yellow!50}{0.001044} & \cellcolor{red!50}{19152} & \cellcolor{green!50}{0.000698} \\
cross-rsdp-128-small & \cellcolor{green!50}{0.000127} & \cellcolor{green!50}{77} & \cellcolor{green!50}{32} & \cellcolor{red!50}{0.006192} & \cellcolor{yellow!50}{10080} & \cellcolor{red!50}{0.004611} \\
cross-rsdp-192-balanced & \cellcolor{green!50}{0.000201} & \cellcolor{green!50}{115} & \cellcolor{green!50}{48} & \cellcolor{orange!50}{0.003711} & \cellcolor{red!50}{28222} & \cellcolor{orange!50}{0.002215} \\
cross-rsdp-192-fast & \cellcolor{green!50}{0.000156} & \cellcolor{green!50}{115} & \cellcolor{green!50}{48} & \cellcolor{yellow!50}{0.002200} & \cellcolor{red!50}{42682} & \cellcolor{yellow!50}{0.001167} \\
cross-rsdp-192-small & \cellcolor{green!50}{0.000129} & \cellcolor{green!50}{115} & \cellcolor{green!50}{48} & \cellcolor{red!50}{0.007809} & \cellcolor{red!50}{23642} & \cellcolor{red!50}{0.005481} \\
cross-rsdp-256-balanced & \cellcolor{green!50}{0.000247} & \cellcolor{green!50}{153} & \cellcolor{green!50}{64} & \cellcolor{orange!50}{0.005708} & \cellcolor{red!50}{51056} & \cellcolor{orange!50}{0.003833} \\
cross-rsdp-256-fast & \cellcolor{green!50}{0.000190} & \cellcolor{green!50}{153} & \cellcolor{green!50}{64} & \cellcolor{orange!50}{0.004123} & \cellcolor{red!50}{76298} & \cellcolor{yellow!50}{0.002273} \\
cross-rsdp-256-small & \cellcolor{green!50}{0.000170} & \cellcolor{green!50}{153} & \cellcolor{green!50}{64} & \cellcolor{red!50}{0.010843} & \cellcolor{red!50}{43592} & \cellcolor{red!50}{0.006649} \\
cross-rsdpg-128-balanced & \cellcolor{green!50}{0.000153} & \cellcolor{green!50}{54} & \cellcolor{green!50}{32} & \cellcolor{yellow!50}{0.001207} & \cellcolor{yellow!50}{9236} & \cellcolor{green!50}{0.000751} \\
cross-rsdpg-128-fast & \cellcolor{green!50}{0.000091} & \cellcolor{green!50}{54} & \cellcolor{green!50}{32} & \cellcolor{green!50}{0.000750} & \cellcolor{orange!50}{12472} & \cellcolor{green!50}{0.000447} \\
cross-rsdpg-128-small & \cellcolor{green!50}{0.000083} & \cellcolor{green!50}{54} & \cellcolor{green!50}{32} & \cellcolor{orange!50}{0.004149} & \cellcolor{yellow!50}{7956} & \cellcolor{orange!50}{0.002552} \\
cross-rsdpg-192-balanced & \cellcolor{green!50}{0.000170} & \cellcolor{green!50}{83} & \cellcolor{green!50}{48} & \cellcolor{yellow!50}{0.001873} & \cellcolor{red!50}{23380} & \cellcolor{yellow!50}{0.001116} \\
cross-rsdpg-192-fast & \cellcolor{green!50}{0.000229} & \cellcolor{green!50}{83} & \cellcolor{green!50}{48} & \cellcolor{yellow!50}{0.001827} & \cellcolor{red!50}{27404} & \cellcolor{green!50}{0.000946} \\
cross-rsdpg-192-small & \cellcolor{green!50}{0.000118} & \cellcolor{green!50}{83} & \cellcolor{green!50}{48} & \cellcolor{red!50}{0.006600} & \cellcolor{red!50}{18188} & \cellcolor{orange!50}{0.003932} \\
cross-rsdpg-256-balanced & \cellcolor{green!50}{0.000211} & \cellcolor{green!50}{106} & \cellcolor{green!50}{64} & \cellcolor{orange!50}{0.003118} & \cellcolor{red!50}{40134} & \cellcolor{yellow!50}{0.001817} \\
cross-rsdpg-256-fast & \cellcolor{green!50}{0.000133} & \cellcolor{green!50}{106} & \cellcolor{green!50}{64} & \cellcolor{yellow!50}{0.002606} & \cellcolor{red!50}{48938} & \cellcolor{yellow!50}{0.001656} \\
cross-rsdpg-256-small & \cellcolor{green!50}{0.000133} & \cellcolor{green!50}{106} & \cellcolor{green!50}{64} & \cellcolor{red!50}{0.009286} & \cellcolor{red!50}{32742} & \cellcolor{red!50}{0.005427} \\
\bottomrule
\end{tabular}
}
  \caption{PQC Signature Schemes: Best and worst in terms of key generation times, public key sizes, private key size, signing time, signatures sizes and verification times.}\label{tab:pqc_signature_schemes}
\end{table}

Similarly, for KEM schemes, Table \ref{tab:pqc_kem_schemes} shows the comparison between BIKE, Classic-McEliece, HQC, ML-KEM, and FrodoKEM in terms of various factors. 
In comparison, Kyber and ML-KEM seem to have better key generation times, smaller public key and private keys, encapsulation times, ciphertext sizes, and decap times.
FrodoKEM seems worst in all factors, while BIKE and Mc-Eliece have better encapsulation times and cipher text sizes.
Among the variations of BIKE, BIKE-L1 has the best encapsulation times and ciphertext sizes.
This shows how the security level, different schemes, has different performance impacts on the KEM schemes.

\begin{table}[!htbp]
\centering
\resizebox{\columnwidth}{!}{
\begin{tabular}{lcccccc}
\toprule
algorithms & KeyGen Times & PK Sizes & SK Sizes & Encap Times & CipherText Sizes & Decap Times \\
\midrule
BIKE-L1 & \cellcolor{yellow!50}{0.000654} & \cellcolor{yellow!50}{1541} & \cellcolor{yellow!50}{5223} & \cellcolor{green!50}{0.000183} & \cellcolor{green!50}{1573} & \cellcolor{red!50}{0.002071} \\
BIKE-L3 & \cellcolor{cyan!50}{0.001559} & \cellcolor{orange!50}{3083} & \cellcolor{orange!50}{10105} & \cellcolor{green!50}{0.000322} & \cellcolor{yellow!50}{3115} & \cellcolor{red!50}{0.005832} \\
BIKE-L5 & \cellcolor{yellow!50}{0.004185} & \cellcolor{red!50}{5122} & \cellcolor{red!50}{16494} & \cellcolor{green!50}{0.000558} & \cellcolor{red!50}{5154} & \cellcolor{red!50}{0.020448} \\
Classic-McEliece-348864 & \cellcolor{red!50}{0.085672} & \cellcolor{red!50}{261120} & \cellcolor{yellow!50}{6492} & \cellcolor{green!50}{0.000189} & \cellcolor{green!50}{96} & \cellcolor{green!50}{0.000124} \\
Classic-McEliece-348864f & \cellcolor{red!50}{0.083439} & \cellcolor{red!50}{261120} & \cellcolor{yellow!50}{6492} & \cellcolor{green!50}{0.000157} & \cellcolor{green!50}{96} & \cellcolor{green!50}{0.000116} \\
Classic-McEliece-460896 & \cellcolor{red!50}{0.287937} & \cellcolor{red!50}{524160} & \cellcolor{orange!50}{13608} & \cellcolor{green!50}{0.000223} & \cellcolor{green!50}{156} & \cellcolor{green!50}{0.000140} \\
Classic-McEliece-460896f & \cellcolor{red!50}{0.171233} & \cellcolor{red!50}{524160} & \cellcolor{orange!50}{13608} & \cellcolor{green!50}{0.000171} & \cellcolor{green!50}{156} & \cellcolor{green!50}{0.000139} \\
Classic-McEliece-6688128 & \cellcolor{red!50}{0.264417} & \cellcolor{red!50}{1044992} & \cellcolor{orange!50}{13932} & \cellcolor{green!50}{0.000451} & \cellcolor{green!50}{208} & \cellcolor{green!50}{0.000163} \\
Classic-McEliece-6688128f & \cellcolor{red!50}{0.251967} & \cellcolor{red!50}{1044992} & \cellcolor{orange!50}{13932} & \cellcolor{green!50}{0.000384} & \cellcolor{green!50}{208} & \cellcolor{green!50}{0.000169} \\
Classic-McEliece-6960119 & \cellcolor{red!50}{0.229219} & \cellcolor{red!50}{1047319} & \cellcolor{orange!50}{13948} & \cellcolor{green!50}{0.000402} & \cellcolor{green!50}{194} & \cellcolor{green!50}{0.000147} \\
Classic-McEliece-6960119f & \cellcolor{red!50}{0.227967} & \cellcolor{red!50}{1047319} & \cellcolor{orange!50}{13948} & \cellcolor{green!50}{0.000335} & \cellcolor{green!50}{194} & \cellcolor{yellow!50}{0.000247} \\
Classic-McEliece-8192128 & \cellcolor{red!50}{0.238020} & \cellcolor{red!50}{1357824} & \cellcolor{orange!50}{14120} & \cellcolor{green!50}{0.000439} & \cellcolor{green!50}{208} & \cellcolor{green!50}{0.000165} \\
Classic-McEliece-8192128f & \cellcolor{red!50}{0.250456} & \cellcolor{red!50}{1357824} & \cellcolor{orange!50}{14120} & \cellcolor{yellow!50}{0.001993} & \cellcolor{green!50}{208} & \cellcolor{yellow!50}{0.000290} \\
HQC-128 & \cellcolor{yellow!50}{0.003869} & \cellcolor{orange!50}{2249} & \cellcolor{yellow!50}{2305} & \cellcolor{red!50}{0.008809} & \cellcolor{orange!50}{4433} & \cellcolor{red!50}{0.014796} \\
HQC-192 & \cellcolor{red!50}{0.012945} & \cellcolor{red!50}{4522} & \cellcolor{yellow!50}{4586} & \cellcolor{red!50}{0.020972} & \cellcolor{red!50}{8978} & \cellcolor{red!50}{0.023882} \\
HQC-256 & \cellcolor{red!50}{0.024194} & \cellcolor{red!50}{7245} & \cellcolor{orange!50}{7317} & \cellcolor{red!50}{0.030119} & \cellcolor{red!50}{14421} & \cellcolor{red!50}{0.041161} \\
Kyber512 & \cellcolor{green!50}{0.000237} & \cellcolor{green!50}{800} & \cellcolor{green!50}{1632} & \cellcolor{green!50}{0.000093} & \cellcolor{green!50}{768} & \cellcolor{green!50}{0.000030} \\
Kyber768 & \cellcolor{green!50}{0.000192} & \cellcolor{green!50}{1184} & \cellcolor{green!50}{2400} & \cellcolor{yellow!50}{0.000670} & \cellcolor{green!50}{1088} & \cellcolor{green!50}{0.000046} \\
Kyber1024 & \cellcolor{green!50}{0.000293} & \cellcolor{green!50}{1568} & \cellcolor{green!50}{3168} & \cellcolor{green!50}{0.000113} & \cellcolor{green!50}{1568} & \cellcolor{green!50}{0.000082} \\
ML-KEM-512 & \cellcolor{green!50}{0.000106} & \cellcolor{green!50}{800} & \cellcolor{green!50}{1632} & \cellcolor{green!50}{0.000049} & \cellcolor{green!50}{768} & \cellcolor{green!50}{0.000079} \\
ML-KEM-768 & \cellcolor{green!50}{0.000154} & \cellcolor{green!50}{1184} & \cellcolor{green!50}{2400} & \cellcolor{green!50}{0.000209} & \cellcolor{green!50}{1088} & \cellcolor{green!50}{0.000040} \\
ML-KEM-1024 & \cellcolor{green!50}{0.000120} & \cellcolor{green!50}{1568} & \cellcolor{green!50}{3168} & \cellcolor{green!50}{0.000095} & \cellcolor{green!50}{1568} & \cellcolor{green!50}{0.000069} \\
sntrup761 & \cellcolor{cyan!50}{0.001088} & \cellcolor{green!50}{1158} & \cellcolor{green!50}{1763} & \cellcolor{green!50}{0.000150} & \cellcolor{green!50}{1039} & \cellcolor{green!50}{0.000088} \\
FrodoKEM-640-AES & \cellcolor{cyan!50}{0.001102} & \cellcolor{red!50}{9616} & \cellcolor{red!50}{19888} & \cellcolor{orange!50}{0.001536} & \cellcolor{red!50}{9720} & \cellcolor{orange!50}{0.001359} \\
FrodoKEM-640-SHAKE & \cellcolor{yellow!50}{0.002303} & \cellcolor{red!50}{9616} & \cellcolor{red!50}{19888} & \cellcolor{red!50}{0.002611} & \cellcolor{red!50}{9720} & \cellcolor{red!50}{0.003413} \\
FrodoKEM-976-AES & \cellcolor{yellow!50}{0.001916} & \cellcolor{red!50}{15632} & \cellcolor{red!50}{31296} & \cellcolor{red!50}{0.002384} & \cellcolor{red!50}{15744} & \cellcolor{orange!50}{0.002292} \\
FrodoKEM-976-SHAKE & \cellcolor{yellow!50}{0.005855} & \cellcolor{red!50}{15632} & \cellcolor{red!50}{31296} & \cellcolor{red!50}{0.005729} & \cellcolor{red!50}{15744} & \cellcolor{red!50}{0.005479} \\
FrodoKEM-1344-AES & \cellcolor{yellow!50}{0.003879} & \cellcolor{red!50}{21520} & \cellcolor{red!50}{43088} & \cellcolor{red!50}{0.004592} & \cellcolor{red!50}{21632} & \cellcolor{red!50}{0.004136} \\
FrodoKEM-1344-SHAKE & \cellcolor{red!50}{0.009928} & \cellcolor{red!50}{21520} & \cellcolor{red!50}{43088} & \cellcolor{red!50}{0.009262} & \cellcolor{red!50}{21632} & \cellcolor{red!50}{0.008229} \\
\bottomrule
\end{tabular}
}
\caption{PQC KEM Schemes: Best and worst in terms of key generation times, public key sizes, private key sizes, encapsulation times, ciphertext size and decapsulation times.}
\label{tab:pqc_kem_schemes}
\end{table}

\subsection{Default Library Results}
Table \ref{tab:sig_details}, presents the default details for various post-quantum signature algorithms, including their NIST security levels and sizes (in bytes) of public keys (PK), secret keys (SK) and signatures (Sig) as provided in the liboqs library. 
The smallest (best) and largest (worst) values for PK, SK, and Sig are colored green and red, respectively, to indicate size efficiency. For example, SPHINCS+-SHA2-128f-simple has the smallest PK and SK, while cross-rsdp-256-fast has the largest Sig.

\begin{table}[!htbp]
\caption{Signature Algorithms Default Details}
\label{tab:sig_details}
\resizebox{\columnwidth}{!}{
\begin{tabular}{lrrrr}
\toprule
Algorithm & Level & PK & SK & Sig \\
\midrule
Dilithium2 & 2 & 1312 & 2528 & 2420 \\
Dilithium3 & 3 & 1952 & 4000 & 3293 \\
Dilithium5 & 5 & 2592 & 4864 & 4595 \\
ML-DSA-44 & 2 & 1312 & 2560 & 2420 \\
ML-DSA-65 & 3 & 1952 & 4032 & 3309 \\
ML-DSA-87 & 5 & 2592 & \cellcolor{red!50}{4896} & 4627 \\
Falcon-512 & 1 & 897 & 1281 & \cellcolor{green!50}{752} \\
Falcon-1024 & 5 & 1793 & 2305 & 1462 \\
Falcon-padded-512 & 1 & 897 & 1281 & 666 \\
Falcon-padded-1024 & 5 & 1793 & 2305 & 1280 \\
SPHINCS+-SHA2-128f-simple & 1 & \cellcolor{green!50}{32} & \cellcolor{green!50}{64} & \cellcolor{red!50}{17088} \\
SPHINCS+-SHA2-128s-simple & 1 & 32 & 64 & 7856 \\
SPHINCS+-SHA2-192f-simple & 3 & 48 & 96 & \cellcolor{red!50}{35664} \\
SPHINCS+-SHA2-192s-simple & 3 & 48 & 96 & 16224 \\
SPHINCS+-SHA2-256f-simple & 5 & 64 & 128 & 49856 \\
SPHINCS+-SHA2-256s-simple & 5 & 64 & 128 & 29792 \\
SPHINCS+-SHAKE-128f-simple & 1 & 32 & 64 & 17088 \\
SPHINCS+-SHAKE-128s-simple & 1 & 32 & 64 & 7856 \\
SPHINCS+-SHAKE-192f-simple & 3 & 48 & 96 & 35664 \\
SPHINCS+-SHAKE-192s-simple & 3 & 48 & 96 & 16224 \\
SPHINCS+-SHAKE-256f-simple & 5 & 64 & 128 & 49856 \\
SPHINCS+-SHAKE-256s-simple & 5 & 64 & 128 & 29792 \\
MAYO-1 & 1 & 1168 & \cellcolor{green!50}{24} & \cellcolor{green!50}{321} \\
MAYO-2 & 1 & \cellcolor{red!50}{5488} & 24 & 180 \\
MAYO-3 & 3 & 2656 & 32 & 577 \\
MAYO-5 & 5 & 5008 & 40 & 838 \\
cross-rsdp-128-balanced & 1 & 77 & 32 & 12912 \\
cross-rsdp-128-fast & 1 & 77 & 32 & 19152 \\
cross-rsdp-128-small & 1 & 77 & 32 & 10080 \\
cross-rsdp-192-balanced & 3 & 115 & 48 & 28222 \\
cross-rsdp-192-fast & 3 & 115 & 48 & 42682 \\
cross-rsdp-192-small & 3 & 115 & 48 & 23642 \\
cross-rsdp-256-balanced & 5 & 153 & 64 & 51056 \\
cross-rsdp-256-fast & 5 & 153 & 64 & \cellcolor{red!50}{76298} \\
cross-rsdp-256-small & 5 & 153 & 64 & 43592 \\
cross-rsdpg-128-balanced & 1 & \cellcolor{green!50}{54} & 32 & 9236 \\
cross-rsdpg-128-fast & 1 & 54 & 32 & 12472 \\
cross-rsdpg-128-small & 1 & 54 & 32 & 7956 \\
cross-rsdpg-192-balanced & 3 & 83 & 48 & 23380 \\
cross-rsdpg-192-fast & 3 & 83 & 48 & 27404 \\
cross-rsdpg-192-small & 3 & 83 & 48 & 18188 \\
cross-rsdpg-256-balanced & 5 & 106 & 64 & 40134 \\
cross-rsdpg-256-fast & 5 & 106 & 64 & 48938 \\
cross-rsdpg-256-small & 5 & 106 & 64 & 32742 \\
\bottomrule
\end{tabular}
}
\begin{tabular}{p{0.9\columnwidth}}
\textbf{Abbreviations:} Level (Claimed NIST Level), PK (Length Public Key in bytes), SK (Length Secret Key in bytes), Sig (Length Signature in bytes) \\
\end{tabular}
\end{table}

This table \ref{tab:kem_details} outlines the default details for post-quantum key encapsulation mechanism (KEM) algorithms, including NIST security levels, IND-CCA security status, and sizes (in bytes) of public keys (PK), secret keys (SK), ciphertexts (CT) and shared secrets (SS). 
The smallest (best) and largest (worst) values for PK, SK, CT, and SS are highlighted in green and red, respectively, to indicate size efficiency. For example, Kyber512 offers the smallest PK and CT, while Classic-McEliece-8192128 has the largest PK.

\begin{table}[!htbp]
\caption{KEM Algorithms Default Details}
\label{tab:kem_details}
\resizebox{\columnwidth}{!}{
\begin{tabular}{lllllll}
\toprule
Algorithm & Level & IND-CCA & PK & SK & CT & SS \\
\midrule
BIKE-L1 & 1 & False & 1541 & 5223 & 1573 & 32 \\
BIKE-L3 & 3 & False & 3083 & 10105 & 3115 & 32 \\
BIKE-L5 & 5 & False & 5122 & \cellcolor{red!50}{16494} & 5154 & 32 \\
Classic-McEliece-348864 & 1 & True & \cellcolor{red!50}{261120} & 6492 & \cellcolor{green!50}{96} & 32 \\
Classic-McEliece-348864f & 1 & True & 261120 & 6492 & 96 & 32 \\
Classic-McEliece-460896 & 3 & True & 524160 & 13608 & 156 & 32 \\
Classic-McEliece-460896f & 3 & True & 524160 & 13608 & 156 & 32 \\
Classic-McEliece-6688128 & 5 & True & 1044992 & 13932 & 208 & 32 \\
Classic-McEliece-6688128f & 5 & True & 1044992 & 13932 & 208 & 32 \\
Classic-McEliece-6960119 & 5 & True & 1047319 & 13948 & 194 & 32 \\
Classic-McEliece-6960119f & 5 & True & 1047319 & 13948 & 194 & 32 \\
Classic-McEliece-8192128 & 5 & True & \cellcolor{red!50}{1357824} & 14120 & 208 & 32 \\
Classic-McEliece-8192128f & 5 & True & 1357824 & 14120 & 208 & 32 \\
HQC-128 & 1 & True & 2249 & 2305 & 4433 & \cellcolor{green!50}{64} \\
HQC-192 & 3 & True & 4522 & 4586 & 8978 & 64 \\
HQC-256 & 5 & True & 7245 & 7317 & \cellcolor{red!50}{14421} & 64 \\
Kyber512 & 1 & True & \cellcolor{green!50}{800} & 1632 & \cellcolor{green!50}{768} & 32 \\
Kyber768 & 3 & True & 1184 & 2400 & 1088 & 32 \\
Kyber1024 & 5 & True & 1568 & 3168 & 1568 & 32 \\
ML-KEM-512 & 1 & True & 800 & 1632 & 768 & 32 \\
ML-KEM-768 & 3 & True & 1184 & 2400 & 1088 & 32 \\
ML-KEM-1024 & 5 & True & 1568 & 3168 & 1568 & 32 \\
sntrup761 & 2 & True & 1158 & 1763 & 1039 & 32 \\
FrodoKEM-640-AES & 1 & True & 9616 & 19888 & 9720 & \cellcolor{red!50}{16} \\
FrodoKEM-640-SHAKE & 1 & True & 9616 & 19888 & 9720 & 16 \\
FrodoKEM-976-AES & 3 & True & 15632 & 31296 & 15744 & 24 \\
FrodoKEM-976-SHAKE & 3 & True & 15632 & 31296 & 15744 & 24 \\
FrodoKEM-1344-AES & 5 & True & 21520 & \cellcolor{red!50}{43088} & 21632 & 32 \\
FrodoKEM-1344-SHAKE & 5 & True & 21520 & 43088 & 21632 & 32 \\
\bottomrule
\end{tabular}
}
\begin{tabular}{p{0.9\columnwidth}}
\textbf{Abbreviations:} Level (Claimed NIST Level), IND-CCA (IND-CCA Security), PK (Length Public Key in bytes), SK (Length Secret Key in bytes), CT (Length Ciphertext in bytes), SS (Length Shared Secret in bytes) \\
\end{tabular}
\end{table}

\end{document}